\documentclass[journal]{IEEEtran}
\usepackage{amssymb}
\usepackage[cmex10]{amsmath}
\usepackage{overpic}
\usepackage{algorithm}
\usepackage{algorithmic}
\usepackage{multirow}
\usepackage{array}
\usepackage{graphicx}
\usepackage{subfigure}
\usepackage{bm}
\usepackage[caption=false,font=normalsize,labelfont=sf,textfont=sf]{subfig}
\usepackage{color}
\usepackage{tikz}
\usepackage{epstopdf}
\usepackage{stfloats}
\usepackage{cite}
\usepackage{tikz}
\usepackage{amsmath}
\usepackage{psfrag}
\usepackage{acronym}
\usepackage{enumerate}
\usetikzlibrary{arrows}
\usetikzlibrary{decorations}

\newcounter{tempEqCounter}

\definecolor{BLUE}{rgb}{0,0,1}

\newtheorem{corollary}{Corollary}
\newtheorem{proposition}{Proposition}
\newtheorem{remark}{Remark}
\newtheorem{lemma}{Lemma}
\newtheorem{definition}{Definition}
\newtheorem{theorem}{Theorem}

\newcommand{\tr}[1]{{\rm tr}\left\{#1\right\}}

\newcommand{\diag}[1]{{\rm diag}\left\{#1\right\}}

\newcommand{\floor}[1]{\left\lfloor#1\right\rfloor}
\newcommand{\rdeg}[1]{\rv{d}_{{\rm A},#1}}

\acrodef{aoa}[AOA]{angle of arrival}
\acrodef{bcrb}[BCRB]{Bayesian Cram\'{e}r-Rao bound}
\acrodef{bp}[BP]{belief propagation}
\acrodef{cl}[CL]{cooperative localization}
\acrodef{crb}[CRB]{Cram\'{e}r-Rao bound}
\acrodef{crlb}[CRLB]{Cram\'{e}r-Rao lower bound}
\acrodef{fim}[FIM]{Fisher information matrix}
\acrodef{mse}[MSE]{mean-squared error}
\acrodef{pdf}[PDF]{probability density function}
\acrodef{pmf}[PMF]{probability mass function}
\acrodef{pll}[PLL]{phase-locked loop}
\acrodef{rbs}[RBS]{reference broadcast synchronization}
\acrodef{rhs}[RHS]{right hand side}
\acrodef{rii}[RII]{ranging information intensity}
\acrodef{rss}[RSS]{received signal strength}
\acrodef{toa}[TOA]{time of arrival}
\acrodef{tpsn}[TPSN]{time synchronization protocol for sensor network}
\acrodef{vmp}[VMP]{variational message passing}
\acrodef{wsn}[WSN]{wireless sensor network}
\acrodef{mse}[MSE]{mean-squared error}

\acrodef{ici}[ICI]{information coupling intensity}
\acrodef{cdi}[CDI]{cooperative dilution intensity}
\acrodef{aseb}[ASEB]{absolute synchronization error bound}
\acrodef{rseb}[RSEB]{relative synchronization error bound}

\usepackage{winsnotation}
\usepackage{bbm}
\newcommand{\fitsize}{\fontsize{8.4pt}{\baselineskip}\selectfont}

\newcommand{\np}{number of equivalent observations for prior distributions}

\begin{document}
\title{Cooperative Network Synchronization: \\ Asymptotic Analysis}

\author{Yifeng~Xiong, \IEEEmembership{Student~Member,~IEEE}, Nan~Wu, \IEEEmembership{Member,~IEEE}, Yuan~Shen, \IEEEmembership{Member,~IEEE}, \\
and Moe~Z.~Win, \IEEEmembership{Fellow,~IEEE}
\thanks{This work was supported in part by the ``National Science Foundation of China (NSFC)'' (Grant No. 61571041, 61421001, 61501279, 91638204) and ``A Foundation for the Author of National Excellent Doctoral Dissertation of P.R. China (FANEDD)'' (Grant No. 201445).}
\thanks{Y. Xiong and N. Wu are with the School of Information and Electronics, Beijing Institute of Technology, Beijing 100081, China (e-mail: yfxiong@bit.edu.cn, wunan@bit.edu.cn).}
\thanks{Y. Shen is with the Department of Electronic Engineering, and Tsinghua National Laboratory for Information Science and Technology, Tsinghua University, Beijing 100084, China (e-mail: shenyuan\_ee@tsinghua.edu.cn).}
\thanks{M.\ Z.\ Win is with the Laboratory for Information and Decision Systems (LIDS), Massachusetts Institute of Technology, Cambridge, MA 02139 USA (e-mail: {moewin@mit.edu}).}
}

\maketitle

\markboth{Accepted by IEEE Transactions on Signal Processing}{Cooperative Network Synchronization: \\ Asymptotic Analysis}

\begin{abstract}
Accurate clock synchronization is required for collaborative operations among nodes across wireless networks. Compared with traditional layer-by-layer methods, cooperative network synchronization techniques lead to significant improvement in performance, efficiency, and robustness. This paper develops a framework for the performance analysis of cooperative network synchronization. We introduce the concepts of \ac{cdi} and relative \ac{cdi} to characterize the interaction between agents, which can be interpreted as properties of a random walk over the network. Our approach enables us to derive closed-form asymptotic expressions of performance limits, relating them to the quality of observations as well as network topology.
\end{abstract}

\begin{IEEEkeywords}
Cooperative network synchronization, \ac{crb}, cooperative dilution intensity (CDI), relative CDI, random walk.
\end{IEEEkeywords}

\section{Introduction}\label{sec:intro}
\IEEEPARstart{N}{etwork synchronization} is a crucial functionality in wireless applications, including geolocation \cite{network_localization,SheWin:J10a, moura_loc,network_exp}, scheduling\cite{scheduling1,scheduling2,scheduling3,schedulingshen}, data fusion\cite{moura_fusion1,varshney_fusion1,moura_fusion2,ZabCon:16}, target tracking \cite{varshney_track1,varshney_track2,BarGioWinCon:J15, BarConGioWin:J14,MeyBraWilHla:J17, PaoGioChiMinMon:08}, and resource utilization \cite{powershen, energyshen, WanLueHua:09, efficient_netloc}. To perform these tasks in a collaborative fashion, nodes are required to operate under a common clock across the network. However, the clocks in nodes suffer from various imperfections caused by both internal and environmental issues, calling for efficient synchronization techniques.

There has been a rich literature on synchronization techniques in wireless networks (WNs) \cite{clock_proceeding,clock_survey,RBS,TPSN}. Traditional methods typically rely on the acyclic structure of the network, among which the most representative ones are \ac{rbs} \cite{RBS} and \ac{tpsn} \cite{TPSN}. These methods are performed in a layer-by-layer manner, requiring high overhead to maintain the acyclic structure and are not robust to link failures. These issues have been addressed by \emph{cooperative synchronization}. Cooperative protocols do not rely on certain hierarchical network structures or special nodes, hence are scalable and insensitive to network topology variations. Existing approaches of this type include consensus-based methods \cite{consensus, consensus_sync1,consensus_sync2} and Bayesian inference methods \cite{mei_leng,EtzMeyHlaSprWym:J17, weijie_yuan}. The convergence of cooperative synchronization methods has been investigated in \cite{consensus_convergence,bp_conv2,conv_fr}.

Understanding the performance of cooperative network synchronization can lead to insights into network deployment and network operation techniques. Performance analysis of network synchronization was pioneered by \cite{lindsey_1}. Early works were influenced by the \ac{pll} structure which were widely used in single-link synchronization. Following this line, a body of literature was devoted to the analysis of network-wide \ac{pll}-based methods \cite{lindsey_2,lindsey_3,lindsey_4}. The issue with this approach is that it focuses on a specific system structure and thus the results therein do not generalize to other systems. Another body of literature, including the seminal work \cite{clock_fundamental}, focused on the feasibility of network synchronization instead of accuracy. Recently, performance limits derived from the information inequality, also known as the Cram\'{e}r-Rao lower bound (CRB), are introduced to address the problem. These performance limits are not restricted to certain system structures or determined by specific algorithm implementations. Thus, they can better reflect the relation between synchronization accuracy and network parameters. However, existing works typically provide only complicated expressions that are not in closed form \cite{coop_crb1,coop_crb2}.

Due to the difference among application scenarios, the network synchronization problem takes various forms, as summarized in \cite{clock_fundamental}. In this paper, we consider two variants of the network synchronization problem, namely \textit{absolute synchronization} and \textit{relative synchronization}. In absolute synchronization, agents are required to be synchronized to a reference clock in reference nodes. In contrast, in relative synchronization there is no reference node, and agents are allowed to reach agreement on any common clock.

In this paper, we develop a novel framework for the performance limit analysis of cooperative network synchronization. Based on this framework, we analyze the asymptotic synchronization performance of large-scale networks for both absolute and relative synchronization. The contributions of this paper are summarized as follows:
\begin{itemize}
\item We derive the performance limits for both absolute and relative cooperative network synchronization, using the \ac{bcrb} and the constrained Cram\'{e}r-Rao Bound, respectively.
\item We propose the concept of cooperative dilution intensity (CDI) to characterize the efficiency of cooperation between an agent and its neighboring nodes for the absolute synchronization scheme, and correspondingly relative \ac{cdi} for the relative synchronization scheme.
\item We propose random walk interpretations of \ac{cdi} and relative \ac{cdi}, which relate these concepts to Markov chains. Using these interpretations, we derive scaling laws for the synchronization performance limits in both dense and extended networks.
\item We analyze the \ac{cdi} and relative \ac{cdi} in infinite lattice networks, finite lattice networks and stochastic networks. Further, we develop asymptotic expressions characterizing the relation between these quantities and the network topology, providing insights into their roles in the scaling laws.
\end{itemize}

The rest of this paper is organized as follows. Section \ref{sec:model} introduces the system model and formulates the network synchronization problem. Based on this model, the performance limits are derived and discussed in Section \ref{sec:limit}. With the help of these expressions, in Section \ref{sec:scaling} we investigate the scaling laws for the performance limits, and in Section \ref{sec:topology} we give explicit asymptotic expressions of \ac{cdi} as well as relative \ac{cdi} for specific network topologies. The analytical results are then verified and illustrated using numerical examples in Section \ref{sec:numerical}, and finally, conclusions are drawn in Section \ref{sec:conclusions}.

\textit{Notations:}
Throughout this paper, $\rv{x}$, $\RV{x}$, $\RM{X}$, and $\RS{X}$ denote random variables, random vectors, random matrices and random sets, respectively; Their realizations are denoted as $x$, $\V{x}$, $\M{X}$, and $\Set{X}$, respectively. The $m$-by-$n$ matrix of zeros (resp. ones) is denoted by $\M{0}_{m\times n}$ (resp. $\M{1}_{m\times n}$). The $m$-dimensional vector of zeros (resp. ones) is denoted by $\V{0}_{m}$ (resp. $\M{1}_{m}$). The $m$-by-$m$ identity matrix is denoted by $\M{I}_{m}$: the subscript is removed when there is no confusion. The indicator function of set $\Set{A}$ is denoted as $\mathbbm{1}_{\Set{A}}(\cdot)$. The round-down function is denoted as $\floor{\cdot}$. The notation $[\cdot]_{i,j}$ denotes the $(i,j)$-th element of its argument; $[\cdot]_{\bar{k}}$ stands for the matrix obtained by deleting the $k$th column and the $k$th row of its argument; $[\cdot]_{r_1:r_2,c_1:c_2}$ denotes a submatrix consists of the $r_1$-th to the $r_2$-th row and the $c_1$-th to the $c_2$-th column of its argument. $\|\V{x}\|_p$ stands for the $l_p$ norm, and denotes the $l_2$ norm when the subscript is omitted. $\mathrm{tr}\{\cdot\}$ denotes the trace of a square matrix. $\M{A}\odot\M{B}$ denotes the Hadamard product between matrix $\M{A}$ and $\M{B}$.

The notation $\mathbb{E}_{\RV{x}}\{\cdot\}$ denotes the expectation with respect to $\RV{x}$, and the subscript is omitted when it is clear from the context. The probability of an event is denoted as $\Prob{\cdot}$. The notation $\nabla_{\V{x}}$ denotes the gradient operator with respect to vector $\V{x}$. The functions $f_{\RV{x}}(\V{x})$, $f_{\RV{x}|\RV{y}}(\V{x}|\V{y})$ and $f_{\RV{x}}(\V{x};\V{\theta})$ denote the \ac{pdf} of $\RV{x}$, the conditional \ac{pdf} of $\RV{x}$ given $\RV{y}$, and the \ac{pdf} of $\RV{x}$ parametrized by $\V{\theta}$, respectively. Some Bachmann-Landau notations used extensively in this paper are summarized as follows.

\vspace{2mm}
\noindent
\begin{tabular}{ll}
    $a(n)\sim b(n)$ & $\lim_{n\rightarrow \infty}a(n)(b(n))^{-1} = 1$ \\
    $a(n)= O(b(n))$ & $\mathop{\lim\sup}_{n\rightarrow \infty} a(n)(b(n))^{-1} < \infty$ \\
    $a(n)=\Omega(b(n))$ & $\mathop{\lim\inf}_{n\rightarrow \infty} a(n)(b(n))^{-1} > 0$ \\
    $a(n)=\Theta(b(n))$ & $a(n)=O(b(n))$ and $a(n)=\Omega(b(n))$ \\
\end{tabular}

\section{System Model}\label{sec:model}
\begin{figure}[t]
    \centering
    \psfrag{xx1}{$\rv{\theta}_1$}
    \psfrag{xx2}{$\rv{\theta}_2$}
    \psfrag{xx3}{$\rv{\theta}_3$}
    \psfrag{xx4}{$\rv{\theta}_4$}
    \includegraphics[width=.38\textwidth]{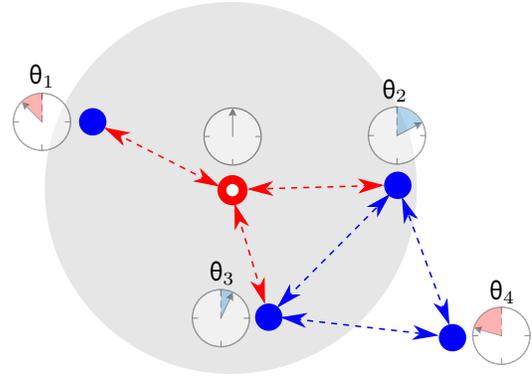}
    \caption{Wireless network synchronization among $N_\mathrm{a} = 4$ agents (blue dots) and $N_\mathrm{r} = 1$ reference node (red circle). Arrows depict communication links.}
    \label{fig:network}
\end{figure}

Consider a network with $N_\mathrm{a}$ nodes with indices comprising a set $\Set{A}=\{1,2,\dotsc,N_\mathrm{a}\}$, each with constant unknown clock offset $\rv{\theta}_i,~i\in\Set{A}$. These nodes are referred to as \textit{agents} hereafter. Additionally, there exists $N_\mathrm{r}$ reference nodes with indices comprising a set $\Set{R}=\{N_\mathrm{a}+1,N_\mathrm{ a}+2,\dotsc,N_\mathrm{a}+N_\mathrm{r}\}$. The network is embedded in $\mathbb{R}^2$, in which node $i$ locates at $\V{p}_i=[p_{\mathrm{x}i}~p_{\mathrm{y}i}]^\mathrm{T}$. Two nodes can communicate with each other if and only if the Euclidean distance between them is no greater than the maximum communication range $R_\mathrm{max}$. We assume that the network is connected, meaning that all agents can communicate with at least one node. We denote the set of all nodes in the communication range of node $i$ as $\Set{N}_i$. An example of such a network is illustrated in Fig. \ref{fig:network}.

The first-order model of the synchronization problem, i.e., only clock offset is considered, is adopted here, which can be expressed as
\begin{equation}\label{fom}
\rv{c}_i(t) = t + \rv{\theta}_i,~i\in\Set{A}
\end{equation}
where $t$ is the reference time and $\rv{c}_i(t)$ is the local clock reading of agent $i$. A more general model including clock skew will be considered in Section \ref{ssec:discussion}.

\begin{figure}[t]
    \centering
    \begin{overpic}[width=.48\textwidth]{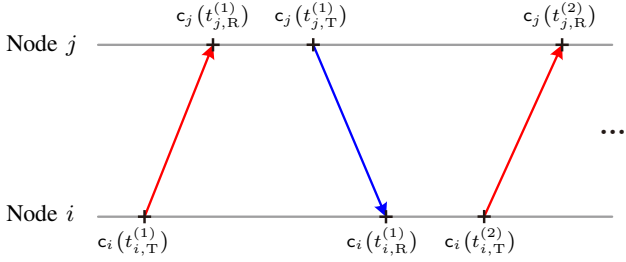}
        \put(15,6){\scriptsize{$\rv{c}_i\big(t_{i,\mathrm{T}}^{(1)}\big)$}}
        \put(53,6){\scriptsize{$\rv{c}_i\big(t_{i,\mathrm{R}}^{(1)}\big)$}}
        \put(68,6){\scriptsize{$\rv{c}_i\big(t_{i,\mathrm{T}}^{(2)}\big)$}}

        \put(27,40.5){\scriptsize{$\rv{c}_j\big(t_{j,\mathrm{R}}^{(1)}\big)$}}
        \put(42,40.5){\scriptsize{$\rv{c}_j\big(t_{j,\mathrm{T}}^{(1)}\big)$}}
        \put(80,40.5){\scriptsize{$\rv{c}_j\big(t_{j,\mathrm{R}}^{(2)}\big)$}}

        \put(1,10){\small {Node $i$}}
        \put(1,36){\small {Node $j$}}
    \end{overpic}
    \caption{Illustration of the two-way timing procedure.}
    \label{fig:clock_model}
\end{figure}

The two-way timing procedure discussed extensively in the literature \cite{coop_crb1,coop_crb2} is illustrated in Fig. \ref{fig:clock_model}. The procedure is started by node $i$, which first sends a message containing its clock reading $\rv{c}_i(t_{i,\mathrm{T}}^{(1)})$ at time $t_{i,\mathrm{T}}^{(1)}$. Node $j$ receives this message at time $t_{j,\mathrm{ R}}^{(1)}$, and replies with a message containing $\rv{c}_j(t_{j,\mathrm{R}}^{(1)})$ and $\rv{c}_j(t_{j,\mathrm{T}}^{(1)})$ at time $t_{j,\mathrm{T}}^{(1)}$, which will be received by node $i$ at time $t_{i,\mathrm{R}}^{(1)}$. When starting the next round, node $i$ will additionally include its clock reading $\rv{c}_i(t_{i,\mathrm{R}}^{(1)})$ in the message. After $N$ such rounds, each node collects $N$ observations $\{\rv{\tau}_{ij}^{(n)}\}_{n=1}^N$ as
\begin{equation}\label{timestamp}
\rv{\tau}_{ij}^{(n)} = \rv{c}_j\big(t_{j,\mathrm{R}}^{(n)}\big) - \rv{c}_i\big(t_{i,\mathrm{T}}^{(n)}\big) + \rv{c}_j\big(t_{j,\mathrm{T}}^{(n)}\big) - \rv{c}_i\big(t_{i,\mathrm{ R}}^{(n)}\big).
\end{equation}
The clock readings in \eqref{timestamp} are related to signal propagation, which are modeled as
\begin{subequations}
\begin{align}
\rv{c}_j\big(t_{j,\mathrm{R}}^{(n)}\big) - \rv{c}_i\big(t_{i,\mathrm{T}}^{(n)}\big) &= \rv{\theta}_j - \rv{\theta}_i + \kappa_{ij} + \rv{\omega}_{n} \\
\rv{c}_j\big(t_{j,\mathrm{T}}^{(n)}\big) - \rv{c}_i\big(t_{i,\mathrm{R}}^{(n)}\big) &= \rv{\theta}_j - \rv{\theta}_i - \kappa_{ji} - \tilde{\rv{\omega}}_{n}
\end{align}
\end{subequations}
where $\kappa_{ij}$ is the deterministic part of message delay (related to processing and signal propagation) which is assume to be symmetric (i.e., $\kappa_{ij}=\kappa_{ji}$), while $\rv{\omega}_{n}$ and $\tilde{\rv{\omega}}_{n}$ denote the stochastic parts (related to signal detection). We assume that $\rv{\omega}_n$ and $\tilde{\rv{\omega}}_{n}$ are independently, identically distributed (i.i.d.) Gaussian random variables, as supported by the measurements presented in \cite{EtzMeyHlaSprWym:J17}. Following previous assumptions, the observations can be rewritten as\footnote{Here, the term $\kappa_{ij}$ is cancelled out by the two-way timing procedure. For systems do not support two-way communication, or clock skews are considered, $\kappa_{ij}$ can be obtained by means of ranging \cite{ranging_1,ranging_2}.}
\begin{equation}\label{timestamp_re}
\rv{\tau}_{ij}^{(n)} = 2(\rv{\theta}_j - \rv{\theta}_i) + \rv{\nu}_n
\end{equation}
where $\rv{\nu}_n=\rv{\omega}_n-\tilde{\rv{\omega}}_{n}$ is a zero-mean Gaussian random variable with variance $\sigma^2$. The joint likelihood function of these observations can thus be obtained as
\begin{equation}\label{jlf}
\begin{aligned}
&f_{\RV{\tau}_{ij}|\rv{\theta}_i,\rv{\theta}_j}(\V{\tau}_{ij}|\theta_i,\theta_j) \\
&\hspace{3mm}=\frac{1}{\left(2\pi\sigma^2\right)^{\frac{N}{2}}}\exp\bigg\{\!-\frac{1}{2\sigma^2}\sum_{n=1}^N\Big[\tau_{ij}^{(n)}-2(\theta_j-\theta_i)\Big]^2\bigg\}
\end{aligned}
\end{equation}
where $\RV{\tau}_{ij}\triangleq[\rv{\tau}_{ij}^{(1)}~\rv{\tau}_{ij}^{(2)}~\dotsc~\rv{\tau}_{ij}^{(N)}]^\mathrm{T}$. To facilitate further derivation, we stack $\RV{\tau}_{ij}$'s in a set $\RS{T}=\left\{\RV{\tau}_{ij}|~\forall i,j,~j\in \Set{N}_i\right\}$.

\textit{Absolute Synchronization: }In absolute synchronization, agents are required to reach an agreement on a specific global clock (e.g., the Global Positioning System). This problem can be formulated as the following estimation problem based on the \ac{mse} minimization criterion
\begin{equation}\label{problem_abs}
\min_{\hat{\RV{\theta}}}\mathbb{E}_{\RS{T},\RV{\theta}}\Big\{\big\|\RV{\theta}-\hat{\RV{\theta}}\big\|^2\Big\}.
\end{equation}
We assume that each node has \textit{a priori} information on their clock offsets, which is modeled as prior distributions on $\left\{\rv{\theta}_i\right\}_{i=1}^{N_\mathrm{a}}$, i.e., $f_{\rv{\theta}_i}(\theta_i)$. The prior distributions are independent across agents.

Based on \eqref{jlf}, the joint distribution of $\RV{\theta}=\left[\rv{\theta}_1~\rv{\theta}_2~\dotsc~\rv{\theta}_{N_\mathrm{a}}\right]^\mathrm{T}$ and $\RS{T}$ can be expressed as
\begin{equation}\label{jlf2}
\begin{aligned}
&f_{\RS{T},\RV{\theta}}(\Set{T},\V{\theta}) \\
&\hspace{1mm}=\prod_{i\in\Set{A}}f_{\rv{\theta}_i}(\theta_i) \prod_{j\in\Set{N}_i}f_{\RV{\tau}_{ij}|\rv{\theta}_i,\rv{\theta}_j}(\V{\tau}_{ij}|\theta_i,\theta_j) \\
&\hspace{1mm}\propto\exp\bigg\{\!-\frac{1}{2\sigma^2}\sum_{i\in\Set{A}}\sum_{j\in\Set{N}_i}\sum_{n=1}^N\Big[\tau_{ij}^{(n)}-2(\theta_j-\theta_i)\Big]^2\bigg\} \\
&\hspace{4mm}\times\prod_{k\in\Set{A}}f_{\rv{\theta}_k}(\theta_k).
\end{aligned}
\end{equation}

\textit{Relative Synchronization:} Some applications only require relative synchronization, where agents are allowed to work under any common clock. In such cases, there is no reference clock, and thus we cannot define prior distributions for agents' clock offsets. Since no prior distribution is involved, we treat the clock offsets in the relative scene as deterministic parameters, denoted by deterministic vector $\V{\theta}$. The corresponding joint likelihood function takes the following form
\begin{equation}\label{rel_jlf}
\begin{aligned}
&f_{\RS{T}}(\Set{T};\V{\theta})\\
&\hspace{3mm}\propto\!\exp\bigg\{\!\!-\!\frac{1}{2\sigma^2}\sum_{i\in\Set{A}}\sum_{j\in\Set{N}_i}\!\sum_{n=1}^N\Big[\tau_{ij}^{(n)}\!-\!2(\theta_j\!-\!\theta_i)\Big]^2\bigg\}.
\end{aligned}
\end{equation}

In order to perform relative synchronization, one can choose an arbitrary agent to be the reference node and perform absolute synchronization. It is shown in \cite{abs_rel} that the error of absolute synchronization can be decomposed as the sum of relative error and transformation error, where the latter is determined by the fixed reference clock. Therefore, given the optimal absolute estimator $\hat{\RV{\theta}}^*$, we can obtain the optimal relative estimator $(\hat{\RV{\theta}}^*,t^*)$ by choosing a reference clock $t^*$, such that the transformation error is minimized as
\begin{equation}\label{problem_rel}
t^*=\mathop{\arg\min}_t\mathbb{E}_{\RS{T}}
\Big\{\big\|\V{\theta}(t)-\hat{\RV{\theta}}^*\big\|^2\Big\}
\end{equation}
where $\V{\theta}(t) = \V{\theta}_0+t$.

\section{Performance Limits}\label{sec:limit}
In this section we derive performance limits for both absolute and relative network synchronization.
\subsection{Absolute Synchronization}
It is well-known that the variance of any unbiased estimators for deterministic parameters are lower bounded by the \ac{crb} \cite{CRB}. For stochastic parameters, the \ac{bcrb} can be used for such tasks. The \ac{bcrb} for the absolute synchronization problem can be obtained using the \ac{fim} defined as
\begin{equation}\label{fim}
\M{J}_{\RV{\theta}}=\mathbb{E}_{\RS{T},\RV{\theta}}\Big\{\left[\nabla_{\RV{\theta}}\ln f_{\RS{T},\RV{\theta}}(\Set{T},\V{\theta})\right]\left[\nabla_{\RV{\theta}}\ln f_{\RS{T},\RV{\theta}}(\Set{T},\V{\theta})\right]^\mathrm{T}\Big\}.
\end{equation}
The following proposition gives the structure of the \ac{fim}.

\begin{proposition}[Structure of the \ac{fim}]\label{prop:structure}
The matrix $\M{J}_{\RV{\theta}}$ takes the form
\begin{equation}\label{decompose}
\M{J}_{\RV{\theta}} = \frac{2N}{\sigma^2}(\M{D}_{\RV{\theta}}^\mathrm{C} + \M{D}_{\RV{\theta}}^\mathrm{R} - \M{A}_{\RV{\theta}}) + \M{\Xi}_{\RV{\theta}}^\mathrm{P}
\end{equation}
where
\begin{equation}\label{defs}
\begin{aligned}
\left[\M{A}_{\RV{\theta}}\right]_{i,j} &= \left\{
                              \begin{array}{ll}
                                1, & \hbox{$j\in\Set{N}_i$} \\
                                0, & \hbox{otherwise}
                              \end{array}
                            \right. \\
\M{D}_{\RV{\theta}}^\mathrm{C} &= \diag{d_{{\rm A},1},d_{{\rm A},2},\dotsc,d_{{\rm A},N_\mathrm{a}}} \\
\M{D}_{\RV{\theta}}^\mathrm{R} &= \diag{d_{{\rm R},1},d_{{\rm R},2},\dotsc,d_{{\rm R},N_\mathrm{a}}} \\
\M{\Xi}_{\RV{\theta}}^\mathrm{P} &= \diag{\xi_{\mathrm{P},1},\xi_{\mathrm{P},2},\dotsc,\xi_{\mathrm{P},N_\mathrm{a}}}
\end{aligned}
\end{equation}
and $d_{{\rm A},i}=\left|\Set{A}\cap\Set{N}_i\right|$ is the number of neighboring agents of agent $i$, $d_{{\rm R},1}=\left|\Set{R}\cap\Set{N}_i\right|$ is the number of neighboring reference nodes of agent $i$. $\M{\Xi}_{\RV{\theta}}^\mathrm{P}$ depicts the \ac{fim} from the \textit{a priori} information of $\RV{\theta}$. The notations C, R, and P correspond to the contributions from cooperation among agents, reference nodes, and prior information, respectively.
\begin{IEEEproof}
See Appendix \ref{sec:proof_structure}.
\end{IEEEproof}
\end{proposition}

With \ac{fim} $\M{J}_{\RV{\theta}}$ given in Proposition \ref{prop:structure}, one can bound the \ac{mse} matrix of estimator $\hat{\RV{\theta}}$ of $\RV{\theta}$ via
\begin{equation}\label{crb}
\mathbb{E}_{\RS{T},\RV{\theta}}\Big\{\big(\RV{\theta}-\hat{\RV{\theta}}\big)
\big(\RV{\theta}-\hat{\RV{\theta}}\big)^\mathrm{T}\Big\}
\succeq \M{J}_{\RV{\theta}}^{-1}.
\end{equation}
Moreover, for any estimator $\hat{\rv{\theta}}_i$ of $\rv{\theta}_i$, its \ac{mse} satisfies \cite{array_geo}
\begin{equation}\label{seb}
\mathbb{E}_{\RS{T},\rv{\theta}_i}\Big\{\big(\rv{\theta}_i-\hat{\rv{\theta}}_i\big)^2\Big\}
\geq \left[\M{J}_{\RV{\theta}}^{-1}\right]_{i,i}.
\end{equation}
Hence, we define the right hand side of \eqref{seb} as a metric of synchronization performance as follows.
\begin{definition}[Absolute Synchronization Error Bound]
The \ac{aseb} of agent $i$ is defined as
$$
s(\rv{\theta}_i)\triangleq\left[\M{J}_{\RV{\theta}}^{-1}\right]_{i,i}.
$$
\end{definition}

The following theorem helps to further understand the physical meaning of the entries of $\M{J}_{\RV{\theta}}^{-1}$.
\begin{theorem}[Structure of Inverse \ac{fim}]\label{thm:decomposition}
When $\M{J}_{\RV{\theta}}$ is invertible, the $(i,j)$-th entry in $\M{J}_{\RV{\theta}}^{-1}$ can be expressed as
\begin{equation}\label{main}
\left[\M{J}_{\RV{\theta}}^{-1}\right]_{i,j} \!=\! \left\{
                                                 \begin{array}{ll}
                                                   \dfrac{1+\Delta_{ii}}{2N\sigma^{-2}\left(d_{{\rm A},i}\!+\!d_{{\rm R},i}\right)+\xi_{\mathrm{P},i}}, & \hbox{$i=j$;} \\
                                                   \dfrac{\Delta_{ij}}{2N\sigma^{-2}\left(d_{{\rm A},j}\!+\!d_{{\rm R},j}\right)+\xi_{\mathrm{P},j}}, & \hbox{$i\neq j$,}
                                                 \end{array}
                                               \right.
\end{equation}
with $\Delta_{ij}\geq 0$ given by
\begin{equation}\label{delta_ij}
\Delta_{ij}\triangleq \sum_{n=1}^{\infty}\left[\M{P}_{\RV{\theta}}^n\right]_{i,j}
\end{equation}
where
\begin{equation}\label{transition_matrix}
\M{P}_{\RV{\theta}}\triangleq \Big(\M{D}_{\RV{\theta}}^\mathrm{C}+\M{D}_{\RV{\theta}}^\mathrm{R}
+\frac{\sigma^2}{2N}\M{\Xi}_{\RV{\theta}}^\mathrm{P}\Big)^{-1}\M{A}_{\RV{\theta}}.
\end{equation}

\begin{IEEEproof}
See Appendix \ref{sec:proof_thm_main}.
\end{IEEEproof}
\end{theorem}

\begin{definition}[Cooperative dilution intensity (CDI)]
The term $\Delta_{ii}$ is referred to as the \ac{cdi} of agent $i$.
\end{definition}
\begin{remark}[Efficiency of cooperation]
From the expression of \ac{aseb} $\M{J}_{\RV{\theta}}^{-1}$, we can see that multiple sources of information contribute to synchronization accuracy. The term $\xi_{\mathrm{P},i}$ accounts for the \textit{a priori} information, while $d_{{\rm A},i}$ and $d_{{\rm R},i}$ correspond to the information from neighboring nodes. The term $(1+\Delta_{ii})^{-1}\in(0,1]$ quantifies the efficiency of cooperation between agent $i$ and its neighbors. Especially, when all neighboring nodes of agent $i$ are reference nodes, we have $\Delta_{ii}=0$. Since $\Delta_{ii}\geq 0$, it can be concluded that the \ac{aseb} reduction from cooperation is not as effective as that directly from reference nodes.
\end{remark}

\begin{remark}[Absolute Synchronizability]
The network is able to perform absolute synchronization only when $\M{J}_{\RV{\theta}}$ is invertible. Here we provide a sufficient condition for absolute synchronizability. Note that $\M{J}_{\RV{\theta}}$ can be rewritten as
$$
\M{J}_{\RV{\theta}} = \Big(\frac{2N}{\sigma^2}\big(\M{D}_{\RV{\theta}}^\mathrm{C}+\M{D}_{\RV{\theta}}^\mathrm{R}\big)
+\M{\Xi}_{\RV{\theta}}^\mathrm{P}\Big)\left(\M{I}-\M{P}_{\RV{\theta}}\right).
$$
To ensure that $\M{J}_{\RV{\theta}}$ is invertible, it suffices to require both $\M{I}-\M{P}_{\RV{\theta}}$ and $2N\sigma^{-2}\big(\M{D}_{\RV{\theta}}^\mathrm{C}+\M{D}_{\RV{\theta}}^\mathrm{R}\big)
+\M{\Xi}_{\RV{\theta}}^\mathrm{P}$ to be invertible. The latter is always invertible since the network is connected. Therefore, absolute synchronizability is guaranteed if the \ac{cdi}s of all agents are finite, i.e., $\Delta_{ii}<\infty,~\forall i$. This condition guarantees the convergence of the matrix power series $\sum_{n=0}^n \M{P}_{\RV{\theta}}^n$, and hence, the invertibility of $\M{I}-\M{P}_{\RV{\theta}}$.
\end{remark}

\begin{proposition}[Node Equivalence]\label{prop:equivalence}
Reference nodes are equivalent to agents with infinite \textit{a priori} information. When agent $k$ has infinite \textit{a priori} information in the sense of $\xi_{\mathrm{P},k} \rightarrow \infty$, we have
\begin{equation}\label{equivalence}
\left[\M{J}_{\RV{\theta}}^{-1}\right]_{\bar{k}} = \left(\left[\M{J}_{\RV{\theta}}\right]_{\bar{k}}\right)^{-1}.
\end{equation}
\begin{IEEEproof}
See Appendix \ref{sec:proof_prop_equivalence}.
\end{IEEEproof}
\end{proposition}

\begin{corollary}[Prior as Virtual Reference Node]\label{coro:virtual}
An agent $i$ with prior information $\xi_{\mathrm{P},i}$ is equivalent to an agent without \textit{a priori} information but connected to an additional reference node providing $N_{\mathrm{p},i} \triangleq \sigma^2\xi_{\mathrm{P},i}(2N)^{-1}$ two-way measurements.
\begin{IEEEproof}
This corollary follows from \eqref{main} and \eqref{equivalence}.
\end{IEEEproof}
\end{corollary}

From Proposition \ref{prop:equivalence} and Corollary \ref{coro:virtual}, we can regard reference nodes and the \textit{a priori} information of agents as additional agents, but with infinite \textit{a priori} information (i.e., knowing that their clock offsets are zeros). In light of this, we denote by $\Set{R}_\mathrm{v}=\left\{N_\mathrm{a}+N_\mathrm{r}+1,N_\mathrm{a}+N_\mathrm{r}+2,\dotsc,2N_\mathrm{a}+N_\mathrm{r}\right\}$ the index set of all virtual reference nodes (priors). A virtual reference node can only communicate with its corresponding agent, i.e., $\Set{N}_i = \{i-N_\mathrm{a}-N_\mathrm{r}|~i\in\Set{R}_\mathrm{v}\}$. We can then construct an extended vector of parameters including the clock offsets of virtual reference nodes as
$$
\tilde{\RV{\theta}}\!=\![\RV{\theta}^\mathrm{T}~\tilde{\rv{\theta}}_{N_\mathrm{a}+1}~\tilde{\rv{\theta}}_{N_\mathrm{a}+2}~\dotsc~
\tilde{\rv{\theta}}_{2N_\mathrm{a}+N_\mathrm{r}}]^\mathrm{T}
$$
and obtain $\M{J}_{\RV{\theta}}^{-1}$ using
$$
\M{J}_{\RV{\theta}}^{-1} = \big[\M{J}_{\tilde{\RV{\theta}}}^{-1}\big]_{1:N_\mathrm{a},1:N_\mathrm{a}}.
$$

The \ac{fim} of $\tilde{\RV{\theta}}$ can be seen as a matrix limit $\M{J}_{\tilde{\RV{\theta}}}=\lim_{\xi_{\rm inf}\rightarrow \infty} \M{J}(\xi_{\rm inf})$, where $\M{J}(\xi_{\rm inf})$ can be partitioned as
\begin{equation}\label{partitioned_FIM}
\M{J}(\xi_{\rm inf})=\left[
                              \begin{array}{ccc}
                                \M{J}_{\RV{\theta}} & -\M{A}_\mathrm{R} & -\M{A}_\mathrm{P}\\
                               -\M{A}_\mathrm{R}^\mathrm{T} & \M{\Xi}_\mathrm{R}(\xi_{\rm inf}) & \M{0}_{N_{\rm r}\times N_{\rm a}}\\
-\M{A}_\mathrm{P}^\mathrm{T} & \M{0}_{N_{\rm a}\times N_{\rm r}} & \M{\Xi}_\mathrm{P}(\xi_{\rm inf})\\
                              \end{array}
                            \right]
\end{equation}
where $\M{\Xi}_\mathrm{R}(\xi_{\rm inf})=\xi_{\rm inf}\M{I}_{N_\mathrm{r}}$ and $\M{\Xi}_\mathrm{P}(\xi_{\rm inf})=\xi_{\rm inf}\M{I}_{N_\mathrm{a}}$ correspond to the ``infinite \textit{a priori} information'' (when taking the limit with respect to $\xi_{\rm inf}$) of reference nodes and virtual reference nodes, respectively. The matrices $\M{A}_\mathrm{R}\in \mathbb{R}^{N_\mathrm{a}\times N_\mathrm{r}}$ and $\M{A}_\mathrm{P}\in \mathbb{R}^{N_\mathrm{a}\times N_\mathrm{a}}$ are given as
\begin{equation}\label{extended_adj1}
\left[\M{A}_\mathrm{R}\right]_{i,j}=\left\{
                                   \begin{array}{ll}
                                     \dfrac{2N}{\sigma^2}, & \hbox{$j\in\Set{N}_i$;} \\
                                     0, & \hbox{otherwise.}
                                   \end{array}
                                 \right.
\end{equation}
and
\begin{equation}\label{extended_adj2}
\left[\M{A}_\mathrm{P}\right]_{i,j}=\left\{
                                   \begin{array}{ll}
                                     \dfrac{2N_{\mathrm{p},i}}{\sigma^2}, & \hbox{$j-i=N_\mathrm{a}+N_\mathrm{r}$;} \\
                                     0, & \hbox{otherwise.}
                                   \end{array}
                                 \right.
\end{equation}
respectively. It can be observed from \eqref{partitioned_FIM}, \eqref{extended_adj1}, and \eqref{extended_adj2} that the matrix $\M{J}_{\tilde{\RV{\theta}}}$ can be regarded as a \ac{fim} of a network with only agents (although some of the agents have infinite \textit{a priori} information).

With Proposition \ref{prop:equivalence} and Corollary \ref{coro:virtual}, we can give the following interpretation of the term $\Delta_{ij}$.
\begin{theorem}[Random Walk Interpretation]\label{thm:random_walk}
The term $\Delta_{ij}$ can be expressed as the following summation
\begin{equation}\label{sum_walk}
\Delta_{ij} = \sum_{n=1}^{\infty}\Prob{\rv{x}_n=j|\rv{x}_0=i}
\end{equation}
where $\Prob{\rv{x}_n=j|\rv{x}_0=i}$ is the $n$-step transition probability of a Markov chain with following one-step transition probability
\begin{equation}\label{random_walk}
\Prob{\rv{x}_k=b|\rv{x}_{k-1}=a}=\left\{
                                   \begin{array}{ll}
                                     -\frac{\left[\M{J}_{\tilde{\RV{\theta}}}\right]_{a,b}}
{\left[\M{J}_{\tilde{\RV{\theta}}}\right]_{a,a}}, & \hbox{$a\neq b$;} \\
                                     0, & \hbox{$a=b,~a\in\Set{A}$;} \\
                                     1, & \hbox{$a=b,~a\in\Set{R}\cup\Set{R}_{\rm v}$.}
                                   \end{array}
                                 \right.
\end{equation}
Especially, $a$ is an absorbing state of the Markov chain if $a\in\Set{R}\cup\Set{R}_\mathrm{v}$.
\begin{IEEEproof}
See Appendix \ref{sec:proof_random_walk}.
\end{IEEEproof}
\end{theorem}

From Theorem \ref{thm:random_walk}, the \ac{cdi} of agent $i$, $\Delta_{ii}$ can be interpreted as the sum of $n$-step return probabilities of the aforementioned Markov chain. Note that the $n$-step transition probabilities are related to the one-step transition probabilities by the recursive Chapman-Kolmogorov equation as
$$
\begin{aligned}
&\Prob{\rv{x}_n=b|\rv{x}_0=a} \\
&= \sum_c\Prob{\rv{x}_n=b|\rv{x}_{n-1}=c}\Prob{\rv{x}_{n-1}=c|\rv{x}_0=a}.
\end{aligned}
$$
Since reference nodes and virtual reference nodes correspond to absorbing states, a random walk starting from state $i$ will never return to its initial state if it reaches such states, and thus the corresponding path will not contribute to $\Delta_{ii}$. Therefore, the way that reference nodes (including virtual ones) provide information about clock offsets can be regarded as ``absorbing'' the random walkers.

\begin{remark}\label{abs_syncable}
According to the Markov chain interpretation of \ac{cdi}, the absolute synchronizability condition ``$\Delta_{ii}<\infty~\forall i$'' can be alternatively stated as that all agents correspond to the transient states of the aforementioned Markov chain. It can also be interpreted as ``for any agent $i$, there exists at least one node with nonzero \textit{a priori} information that can be reached from $i$ in finite steps''.
\end{remark}

\subsection{Relative Synchronization}
Using \eqref{rel_jlf} and following a similar argument as used in Proposition \ref{prop:structure}, one can see that in the relative synchronization scenario, the \ac{fim} $\M{J}_{\V{\theta}}$ is given by
$$
\M{J}_{\V{\theta}}=\frac{2N}{\sigma^2}(\M{D}_{\V{\theta}}^\mathrm{C} - \M{A}_{\V{\theta}}).
$$
The relative \ac{fim} $\M{J}_{\V{\theta}}$ is not invertible since $\M{J}_{\V{\theta}}\V{1}_{N_\mathrm{a}}=\V{0}_{N_\mathrm{a}}$. Nevertheless, it has been shown in \cite{abs_rel} that for any relative estimator $(\hat{\RV{\theta}},t)$, its relative \ac{mse} can be lower bounded using the constrained \ac{crb}, namely the Moore-Penrose pseudo-inverse of the \ac{fim} as
$$
\mathbb{E}_{\RS{T}}\Big\{\big\|\V{\theta}(t)-\hat{\RV{\theta}}\big\|^2\Big\}\geq \mathrm{tr}\big\{\M{J}_{\V{\theta}}^\dagger\big\}.
$$

Similar to the absolute case, in relative synchronization the matrix $\M{J}_{\V{\theta}}^\dagger$ also admits a Markov chain interpretation. In relative synchronization, the Markov chain of interest is characterized by the one-step transition matrix $\M{P}_{\V{\theta}}=(\M{D}_{\V{\theta}}^\mathrm{C})^{-1}\M{A}_{\V{\theta}}$. Here we denote by $p_{ij}^{(n)}$ the $n$-step transition probability from state $i$ to $j$.

\begin{theorem}[Relative \ac{mse} Lower Bound]\label{thm:rel}
For a connected network performing relative synchronization, the relative \ac{mse} bound can be expressed as
\begin{equation}\label{rel_thm}
\mathrm{tr}\big\{\M{J}_{\V{\theta}}^\dagger\big\} = \sum_{i=1}^{N_\mathrm{a}}\frac{1+\tilde{\Delta}_{ii}}{2N\sigma^{-2}d_{{\rm A},i}}
\end{equation}
with $\tilde{\Delta}_{ii}$ given by
\begin{equation}\label{rel_cdi}
\tilde{\Delta}_{ii}\triangleq\sum_{n=1}^\infty \bigg\{\left[\M{P}_{\V{\theta}}^n\right]_{i,i}-\frac{1}{N_\mathrm{a}}\sum_{j=1}^{N_\mathrm{a}}\left[\M{P}_{\V{\theta}}^n\right]_{j,i}\bigg\}.
\end{equation}
\begin{IEEEproof}
See Appendix \ref{sec:proof_rel}.
\end{IEEEproof}
\end{theorem}
\begin{definition}[Relative Synchronization Error Bound]
The \ac{rseb} is defined as
$$
\bar{s}(\V{\theta}) = \frac{1}{N_\mathrm{a}}\mathrm{tr}\big\{\M{J}_{\V{\theta}}^\dagger\big\}.
$$
\end{definition}

It can be seen that \eqref{rel_thm} takes a similar form as \eqref{main}, and the term $\tilde{\Delta}_{ii}$ plays a similar role as the term $\Delta_{ii}$ does in the absolute case. Thus we make the following definition:
\begin{definition}[Relative \ac{cdi}]
The term $\tilde{\Delta}_{ii}$ is referred to as the relative \ac{cdi} of agent $i$.
\end{definition}

According to the properties of the equilibrium distribution of reversible Markov chains, for any agent $j$ in a connected network, the term $\left[\M{P}_{\V{\theta}}^n\right]_{j,i}$ tends to $d_{{\rm A},i}(\sum_{k=1}^{N_\mathrm{a}}d_{{\rm A},k})^{-1}$ as $n\rightarrow \infty$. Hence the series in \eqref{rel_cdi} is guaranteed to converge, and thus connected networks with finite number of agents are always synchronizable in the relative sense.

\subsection{Impact of Clock Skews}\label{ssec:discussion}
Now we consider a general clock model taking clock skews into account
\begin{equation}
\rv{c}_i(t) = \rv{\alpha}_i t + \rv{\theta}_i
\end{equation}
where $\rv{\alpha}_i$ takes positive values around $1$. We consider the case where the clock offsets are random but known in advance. Taking expectation over both sides of \eqref{crb} with respect to the clock skews, we have
\begin{equation}\label{mcb}
\mathbb{E}_{\RS{T},\RV{\theta},\RV{\alpha}}\Big\{\big(\RV{\theta}-\hat{\RV{\theta}}\big)\big(\RV{\theta}-\hat{\RV{\theta}}\big)^\mathrm{T}\Big\}\succeq \mathbb{E}_{\RV{\alpha}}\{\RM{J}_{\RV{\theta}}^{-1}\}.
\end{equation}
where $\RV{\alpha}=[\rv{\alpha}_1~\rv{\alpha}_2~\dotsc~\rv{\alpha}_{N_\mathrm{a}}]^\mathrm{T}$. Using the same two-way timing protocol as described in Section \ref{sec:model}, the \ac{fim} can be expressed as
\begin{equation}\label{FIM_skew}
\left[\RM{J}_{\RV{\theta}}\right]_{i,j} =
\left\{
    \begin{array}{ll}
        \sum_{j\in\Set{N}_i} 2N\rv{\alpha}_i^{-2}\sigma^{-2}, & \hbox{$i=j$;} \\
        -2N\rv{\alpha}_i^{-1}\rv{\alpha}_j^{-1}\sigma^{-2}, & \hbox{otherwise.}
    \end{array}
\right.
\end{equation}
Note that the \ac{fim} is now denoted as $\RM{J}_{\RV{\theta}}$ since it is a random matrix. Denoting the \ac{fim} corresponding to the case without clock skews (i.e., $\alpha_i=1~\forall i\in\Set{A}$) as $\M{J}_{\RV{\theta}}$, we have
\begin{equation}\label{FIM_skew2}
\RM{J}_{\RV{\theta}} = \RM{B}^{-1}\M{J}_{\RV{\theta}}\RM{B}^{-1},
\end{equation}
where $\RM{B} = \diag{\rv{\alpha}_1,\rv{\alpha}_2,\dotsc,\rv{\alpha}_{N_\mathrm{a}}}$. The inverse of $\RM{J}_{\RV{\theta}}$ can thus be calculated as
\begin{equation}\label{FIM_skew2_inv}
\RM{J}_{\RV{\theta}}^{-1} = \RM{B}\M{J}_{\RV{\theta}}^{-1}\RM{B}
\end{equation}
and the Moore-Penrose pseudo-inverse for the relative synchronization scenario is given by
\begin{equation}\label{FIM_skew2_pinv}
\RM{J}_{\RV{\theta}}^{\dagger} = \RM{C}_\mathrm{s}\RM{B}\M{J}_{\RV{\theta}}^\dagger\RM{B}\RM{C}_\mathrm{s}
\end{equation}
where $\RM{C}_\mathrm{s} = \M{I}-\|\RM{B}\V{1}_{N_\mathrm{a}}\|^{-2}\RM{B}\V{1}_{N_\mathrm{a}}\V{1}_{N_\mathrm{a}}^\mathrm{T}\RM{B}^\mathrm{T}$. Assuming that all clock skews have mean value $1$, we have
\begin{equation}
\mathbb{E}_{\RV{\alpha}}\{\RM{J}_{\RV{\theta}}^{-1}\}\succeq \M{J}_{\RV{\theta}}^{-1}
\end{equation}
due to the Jensen's inequality. This implies that introducing clock skews into the network always leads to performance degradation, which agrees with intuition.

From \eqref{FIM_skew2_inv} and \eqref{FIM_skew2_pinv} we can see that the difference between $\RM{J}_{\RV{\theta}}$ and $\M{J}_{\RV{\theta}}$ resides in the matrix $\RM{B}$, which is not related to the network topology. Therefore, from a network-level perspective, we can continue our discussion on the case without clock skews, and the results can then be easily extended to the case with known clock skews.

\section{Scaling Laws}\label{sec:scaling}
In this section, we investigate the scaling laws for proposed performance limits for both absolute and relative synchronization. Scaling laws characterize the asymptotic performance in large networks, and provide insights into the nature of the network synchronization problem.

We consider two types of random networks that are extensively used in the modelling of realistic wireless networks, namely \textit{extended networks} and \textit{dense networks} \cite{coop_fundamental}. In both types of networks, agents are modelled as instances of a binomial point process with intensity $\lambda_\mathrm{a}=N_\mathrm{a}|\Set{R}_\mathrm{net}|^{-1}$ on a region $\Set{R}_\mathrm{net}$. As $N_\mathrm{a}$ grows, the intensity $\lambda_\mathrm{a}$ increases proportionally in dense networks, and remains constant in extended networks.

For simplicity of derivation, in this section we assume that there is no reference node, but agents can have \textit{a priori} information of their clock offsets. This assumption does not influence the generality of the results due to the equivalence between reference nodes and \textit{a priori} information.

\subsection{Extended Networks}
In extended networks, the distribution of neighboring agent number $d_{{\rm A},i}$ for an agent does not change as $N_\mathrm{a}$ increases. Therefore, from \eqref{main} and \eqref{rel_thm} we see that the expected \ac{aseb} scales proportionally to the expected \ac{cdi}, while the expected \ac{rseb} scales proportionally to the expected relative \ac{cdi}, respectively.

\begin{proposition}\label{prop:extended_rel1}
In extended networks performing relative synchronization, $\mathbb{E}\big\{\bar{\rv{s}}(\V{\theta})\big\}\rightarrow \infty$ as $N_\mathrm{a}\rightarrow \infty$.
\begin{IEEEproof}
See Appendix \ref{sec:proof_extended_rel1}.
\end{IEEEproof}
\end{proposition}

Proposition \ref{prop:extended_rel1} implies that relative synchronization is impossible in infinitely large networks. This agrees with the intuition that agreement on a common clock cannot be achieved in infinitely large networks.

For networks performing absolute synchronization, we have a slightly different result as follows.
\begin{proposition}\label{prop:extended_abs1}
In extended networks performing absolute synchronization, if the amount of \textit{a priori} information of any agent is no greater than $\xi_{\max}$, i.e., $N_{\mathrm{p},i}\leq \sigma^2\xi_{\max}(2N)^{-1}$, then as $N_\mathrm{a}\rightarrow \infty$, we have $\E{\rv{s}(\rv{\theta}_i)}=\Omega(1)$.
\begin{IEEEproof}
See Appendix \ref{sec:proof_extended_abs1}.
\end{IEEEproof}
\end{proposition}

Proposition \ref{prop:extended_abs1} states that increasing the network area is not beneficial for extended networks performing absolute synchronization in an asymptotic regime. This reserves the possibility that the expected average \ac{aseb} remains finite as $N_\mathrm{a}\rightarrow \infty$ under certain conditions, which is the main difference from the relative case. In the following proposition, we present one of such conditions.

\begin{proposition}\label{prop:extended_abs2}
Assume that agents having \textit{a priori} information constitute a binomial point process with fixed intensity $\lambda_\mathrm{ap}$, and the amount of \textit{a priori} information is no less than $\xi_{\min}$, i.e., $N_{\mathrm{p},i}\geq \sigma^2\xi_{\min}(2N)^{-1}$. If the conditions in Proposition \ref{prop:extended_abs1} hold, $N_\mathrm{a}\rightarrow \infty$, we have $\mathbb{E}\big\{N_\mathrm{a}^{-1}\sum_{i=1}^{N_\mathrm{a}}\rv{s}(\rv{\theta}_i)\big\}=\Theta(1)$.
\begin{IEEEproof}
See Appendix \ref{sec:proof_extended_abs2}.
\end{IEEEproof}
\end{proposition}

Proposition \ref{prop:extended_abs2} is basically a direct application of Remark \ref{abs_syncable}. By assumption, agents with \textit{a priori} information are randomly located. Therefore, a random walk from any other agent can reach one of these agents in finite steps, meaning that all agents can get access to some information about their clock offsets.

\subsection{Dense Networks}
In dense networks, the expected number of neighboring agents increases proportionally to $N_\mathrm{a}$. Hence, it can be seen from \eqref{main} and \eqref{rel_thm} that both expected \ac{aseb} and expected \ac{rseb} scale as $\Theta(N_\mathrm{a}^{-1})$, as long as \ac{cdi} and relative \ac{cdi} are bounded from above as $N_\mathrm{a}\rightarrow \infty$.
\begin{proposition}\label{prop:dense_rel}
In dense networks performing relative synchronization, the expected \ac{rseb} $\mathbb{E}\big\{\bar{\rv{s}}(\V{\theta})\big\}$ scales as $\Theta(N_\mathrm{a}^{-1})$.
\begin{IEEEproof}
See Appendix \ref{sec:proof_dense_rel}.
\end{IEEEproof}
\end{proposition}

\begin{proposition}\label{prop:dense_abs}
In dense networks performing absolute synchronization, assume that if an agent has \textit{a priori} information, the amount of \textit{a priori} information is no less than $\xi_{\min}$. Then the expected average \ac{aseb} $\mathbb{E}\big\{N_\mathrm{a}^{-1}\sum_{i=1}^{N_\mathrm{a}}\rv{s}(\rv{\theta}_i)\big\}$ scales as $\Theta(N_\mathrm{a}^{-1})$ if one of the following conditions holds:
\begin{enumerate}
\item Agents with \textit{a priori} information constitute a binomial point process with fixed intensity;
\item All agents in a certain region $\Set{R}$ with fixed area have \textit{a priori} information.
\end{enumerate}
\begin{IEEEproof}
See Appendix \ref{sec:proof_dense_abs}.
\end{IEEEproof}
\end{proposition}

Note that the second condition in Proposition \ref{prop:dense_abs} is equivalent to the case where there exist constant number of reference nodes in the network.

\section{Analysis on \ac{cdi} and Relative \ac{cdi}}
\label{sec:topology}
The concepts of \ac{cdi} and relative \ac{cdi} play important roles in the scaling laws for synchronization performance. In networks performing absolute synchronization, the quantity $(1+\mathrm{\ac{cdi}})^{-1}$ also characterizes the efficiency of cooperation between an agent and other agents. Unfortunately, calculation of \ac{cdi} or relative \ac{cdi} is generally intractable since the $n$-step transition probabilities of Markov chains have no closed-form expressions in general.

In this section, we develop some asymptotic expressions of \ac{cdi} as well as relative \ac{cdi} for specific network topologies. Unless otherwise stated, we make following assumptions in the derivation henceforth:
\begin{itemize}
\item There is no explicit reference node in the network, i.e., $d_{{\rm R},i}= 0~\forall i\in\Set{A}$. According to Corollary \ref{coro:virtual}, the \textit{a priori} information of agents can be considered as virtual reference nodes.
\item For networks performing absolute synchronization, all agents have the same amount of \textit{a priori} information, i.e., $\xi_{\mathrm{P},i} = \xi_\mathrm{P}$ and $N_{\mathrm{p},i} = N_\mathrm{p}$ for all $i\in\Set{A}$.
\end{itemize}

\subsection{Infinite Lattice Networks}\label{ssec:inf_lattice}
\begin{figure}[t]
    \centering
    \psfrag{xx1}{Agent $i$}
    \psfrag{k}{$R_{\max}$}
    \psfrag{xx3}{$\Set{N}_i$}
    \includegraphics[width=.36\textwidth]{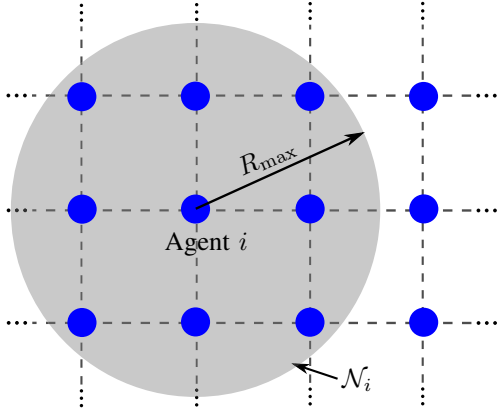}
    \caption{Agents shown in blue circles form an instance of infinite lattice networks. The agents in the gray circle (except agent $i$) consist of the neighborhood $\Set{N}_i$ with radius $R_\mathrm{max}$ of agent $i$.}
    \label{fig:infinite_lattice}
\end{figure}

An infinite lattice network is illustrated in Fig. \ref{fig:infinite_lattice}.
In this kind of networks, there are infinite number of agents, whose positions cover all lattice points (points with integer coordinates) in the space $\mathbb{R}^2$. Under previous assumptions, \eqref{transition_matrix} can be rewritten as
\begin{equation}\label{p_lattice}
\M{P}_{\RV{\theta},\mathrm{IL}} = \frac{1}{\bar{d}+N_\mathrm{p}}\M{A}_{\RV{\theta},\mathrm{IL}}
\end{equation}
where $\bar{d}$, the number of neighboring agents of an agent, is identical for all agents. Here the subscript $\rm IL$ denotes infinite lattice networks, and the definition of the corresponding quantities are the same as those without this subscript. Note that in light of Proposition \ref{prop:extended_rel1}, in infinite lattice networks, the relative \ac{cdi} does not exist, hence in this subsection we focus on the analysis on \ac{cdi}.

According to \eqref{delta_ij} and using the random walk interpretation in Theorem \ref{thm:random_walk}, $\Delta_{ij}$ can be expressed as
\begin{equation}\label{numerical}
\Delta_{ij} = \sum_{n=1}^{\infty} p_{ij}^{(n)}\left(\frac{\bar{d}}{\bar{d}+N_\mathrm{p}}\right)^n
\end{equation}
where $p_{ij}^{(n)}\triangleq \Prob{\rv{x}_n=j|\rv{x}_0=i}$ is the $n$-step transition probability of a Markov chain with following one-step transition probability
\begin{equation}\label{mc_2}
\Prob{\rv{x}_k=j|\rv{x}_{k-1}=i} = \frac{\mathbbm{1}_{\Set{N}_j}(i)}{\bar{d}}.
\end{equation}

With the help of previous results, we can derive an approximated expression for $\Delta_{ij}$. Replacing the states in \eqref{mc_2} as the positions of agents, we obtain
\begin{equation}\label{replace_state}
\Prob{\RV{x}_k=\V{p}_j|\RV{x}_{k-1}=\V{p}_i} = \frac{\mathbbm{1}_{\Set{N}_j}(i)}{\bar{d}}.
\end{equation}
Thus we can introduce an auxiliary stochastic process $\RV{y}_k$ which is strictly stationary, with the following time-invariant distribution
$$
\Prob{\RV{y}_k=\V{y}}=\frac{1}{\bar{d}}\cdot\mathbbm{1}_{(0,R_\mathrm{max}]}(\left\|\V{y}\right\|) \mathbbm{1}_{\mathbb{Z}^2}(\V{y})
$$
so that the states in \eqref{replace_state} can be expressed alternatively as
$$
\RV{x}_k = \RV{x}_{k-1} + \RV{y}_k
$$
and hence
\begin{equation}\label{summation}
\RV{x}_k = \RV{x}_0 + \sum_{n=1}^k \RV{y}_n.
\end{equation}
The \ac{pmf} of $\RV{x}_k$ is the result of $k$ self-convolutions of $\Prob{\RV{y}_n=\V{y}}$, and thus the term $p_{ij}^{(n)}$ can be rewritten as $\Prob{\RV{x}_n=\V{p}_j|\RV{x}_0=\V{p}_i}$.

Now note that the summation in \eqref{summation} is a sum over i.i.d. random variables. By application of the local central limit theorem \cite{localCLT}, we can approximate the conditional \ac{pmf} of $\RV{x}_k$ as Gaussian (on lattice points) so that
\begin{equation}\label{approximation_gaussian}
\begin{aligned}
\Prob{\RV{x}_k=\V{x}|\RV{x}_0=\V{x}_0} &= \frac{1}{2\pi k\sigma_\mathrm{R}^2}\exp\left\{-\frac{1}{k\sigma_\mathrm{R}^2}\|\V{x}-\V{x}_0\|^2\right\} \\
&\hspace{5mm}+ \frac{1}{k}E_1(k,\|\V{x}-\V{x}_0\|)
\end{aligned}
\end{equation}
where $\sigma_\mathrm{R}^2$ is chosen such that
$$
\mathbb{E}_{\RV{x}_1|\RV{x}_0}\Big\{\left.\left(\RV{x}_1-\RV{x}_0\right)\left(\RV{x}_1-\RV{x}_0\right)^\mathrm{T}\right|\RV{x}_0\Big\} = \sigma_\mathrm{R}^2 \M{I}_2,
$$
and $E_1(k,\|\V{x}-\V{x}_0\|)$ is an error term which tends to zero as $k\rightarrow \infty$ for all $\V{x}$. From \eqref{approximation_gaussian} we have the following estimate on the order of the \ac{cdi} $\Delta_{ii}$.
\begin{theorem}[Asymptotic \ac{cdi}]\label{thm:ici}
The \ac{cdi} $\Delta_{ii}$ of infinite lattice networks has the following asymptotic behavior
\begin{equation}\label{delta_ii_approx}
\Delta_{ii} \sim \frac{2}{\bar{d}}\ln\left(1+\frac{\bar{d}}{N_\mathrm{p}}\right)
\end{equation}
as $R_\mathrm{max}\rightarrow \infty$.
\begin{IEEEproof}
See Appendix \ref{sec:proof_ici}.
\end{IEEEproof}
\end{theorem}

\begin{remark}
Note that in infinite lattice networks, $\bar{d}$ is a function of $R_{\max}$ given by \cite{gauss_circle}
$$
\bar{d} = 1+4\floor{R_{\max}}+4\sum_{n=1}^{\floor{R_{\max}}}\floor{\sqrt{R_{\max}^2-n^2}}.
$$
Increasing $\bar{d}$ means that more information from the neighboring nodes can be obtained. Theorem \ref{thm:ici} indicates that as $\bar{d}$ increases, the \ac{cdi} of an arbitrary agent drops, which could be interpreted as ``the neighbors of the agent become more like reference nodes''. It can also be seen from \eqref{delta_ii_approx} that $\Delta_{ii}$ is positively related with $N_\mathrm{p}$, implying that the degree to which agents behave like reference nodes is not determined by the absolute amount of \textit{a priori} information, but depends on the number of equivalent observations for \textit{a priori} information.
\end{remark}

\subsection{Finite Lattice Networks}\label{ssec:fin_lattice}
Finite lattice networks are those with finite number of nodes located on lattice points. The following proposition indicates that for an agents $i$ in a finite lattice network, the \ac{cdi} $\Delta_{ii}$ is never less than that of infinite lattice networks (as long as some technical conditions are satisfied). In the following discussion, subscript $\rm L$ denotes the quantities in a finite lattice network.
\begin{proposition}\label{prop:fin_coupling}
Consider an infinite lattice network $\Set{G}_\mathrm{IL}$ with certain $R_\mathrm{max}$ such that the network is connected. For any finite lattice network $\Set{G}_\mathrm{L}$ obtained by partitioning $\Set{G}_\mathrm{IL}$ with non-empty interior area $\Set{I}$,\footnote{``Interior area'' stands for the set of points with distance at least $R_\mathrm{max}$ to the edge of the network.} the corresponding \ac{cdi} $\Delta_{ii}~\forall i\in\Set{I}$ is no less than that of $\Set{G}_\mathrm{IL}$, if the following assumptions hold:
\begin{enumerate}[A)]
\item \label{as1}The shortest paths between an agent $i\in\Set{I}$ and another agent $j\in\Set{G}_\mathrm{L}$ always reside in $\Set{G}_\mathrm{L}$;
\item \label{as2}For any $i\in\Set{I}$ and any agent $j$ in the edge\footnote{The set of point in a finite lattice network but not in the interior area is called the ``edge''.} (denoted by $j\in\Set{E}$),
$$
\frac{\sum_{k\in\Set{N}_{j,\mathrm{L}}}p_{ki,\mathrm{IL}}^{(n)}}{\left|\Set{N}_{j,\mathrm{L}}\right|}\geq \frac{\sum_{k\in\Set{N}_{j,\mathrm{IL}}\backslash\Set{N}_{j,\mathrm{L}}}p_{ki,\mathrm{IL}}^{(n)}}{\left|\Set{N}_{j,\mathrm{IL}}\backslash\Set{N}_{j,\mathrm{L}}\right|}~\forall n\in\mathbb{Z}_+.
$$
\end{enumerate}

\end{proposition}
\begin{IEEEproof}
See Appendix \ref{sec:proof:fin}.
\end{IEEEproof}

Assumption \ref{as1} implies that $\Set{G}_\mathrm{L}$ should have a convex boundary. Assumption \ref{as2} can be intuitively interpreted as ``in the neighborhood of an agent $j\in\Set{E}$, it is easier for the agents in $\Set{E}$ to reach $\Set{I}$ compared to those outside $\Set{G}_\mathrm{L}$'', which holds for most convex sets. In addition, the number of agents in the interior area dominates as $\Set{G}_\mathrm{L}$ extends. Therefore, for most large finite lattice networks having convex boundaries, the average \ac{cdi} is no less than the \ac{cdi} in infinite lattice networks.

With Proposition \ref{prop:fin_coupling}, we can obtain the following result for the relative \ac{cdi} in finite lattice networks.
\begin{proposition}\label{prop:rel_cdi}
Assume that the assumptions in Proposition \ref{prop:fin_coupling} hold, and as the network expands, we have $|\Set{E}||\Set{I}|^{-1}\rightarrow 0$. Then the average relative \ac{cdi} in finite lattice networks has the following asymptotic behavior as $N_\mathrm{a}\rightarrow\infty$
\begin{equation}\label{rel_cdi1}
\frac{1}{N_\mathrm{a}}\sum_{i=1}^{N_\mathrm{a}}\tilde{\Delta}_{ii} = O(\bar{d}^{-1}\ln N_\mathrm{a})
\end{equation}
where $\bar{d}$ is the average number of neighboring agents for an agent in the network.
\end{proposition}
\begin{IEEEproof}
See Appendix \ref{sec:proof_rel_cdi}.
\end{IEEEproof}
\begin{remark}
Proposition \ref{prop:rel_cdi} implies that, if the communication range $R_{\max}$ does not change, the average relative \ac{cdi} will increase logarithmically (and thus unboundedly) as the network expands. This is similar to the scaling law in extended networks where the relative synchronization error is unbounded as $N_\mathrm{a}\rightarrow \infty$.
\end{remark}

\subsection{Stochastic Networks}\label{ssec:inf_sto}
A stochastic network can be modeled as a network with $N_\mathrm{a}$ agents, distributed uniformly on $[0,1]^2$ \cite{boyd2005mixing}. In this paper, we consider the stochastic network defined on $[0,B]^2,~B\in\mathbb{R}_+$ with constant node intensity $\lambda_\mathrm{s}$ such that $N_\mathrm{a}=\lambda_\mathrm{s} B^2$, and focus on the limiting case as $B\rightarrow \infty$. These agents constitute a binomial point process, which tends to a homogeneous Poisson point process (PPP) \cite{Kin:B93} as $B\rightarrow \infty$ while $\lambda_\mathrm{s}$ remains constant. PPPs are widely used in the modeling of wireless networks \cite{WinPinGioChiShe:06, win2007mathematical,msg_2, HaeAndBacDouFra:09, DarConBurVer:07,RabConWin:J15,rgg2,msg_3,ElSSulAloWin:J17,msg_4,outage}.

In a stochastic network, we are interested in the performance averaged over all possible topologies. Specifically, the expected \ac{cdi} is of interest
\begin{equation}\label{stochastic_cdi}
\mathbb{E}\bigg\{\frac{1}{N_\mathrm{a}}\sum_{i=1}^{N_\mathrm{a}} \Delta_{ii}\bigg\} = \mathbb{E}\bigg\{\frac{1}{N_\mathrm{a}}\tr{\RM{L}_{\RV{\theta},\mathrm{S}}}\bigg\}-1
\end{equation}
with
$$
\RM{L}_{\RV{\theta},\mathrm{S}}\triangleq \left(\M{I}-\RM{P}_{\RV{\theta},\mathrm{S}}\right)^{-1}
$$
where $\RM{P}_{\RV{\theta},\mathrm{S}}$ can be expressed as follows according to \eqref{transition_matrix}
$$
\RM{P}_{\RV{\theta},\mathrm{S}} = \left(\RM{D}_{\RV{\theta},\mathrm{S}}+N_\mathrm{p}\M{I}\right)^{-1} \RM{A}_{\RV{\theta},\mathrm{S}}
$$
and the subscript $\rm S$ denotes stochastic networks. Note that $\RM{L}_{\RV{\theta},\mathrm{S}}$, $\RM{P}_{\RV{\theta},\mathrm{S}}$, $\RM{D}_{\RV{\theta},\mathrm{S}}$, and $\RM{A}_{\RV{\theta},\mathrm{S}}$ are random matrices in stochastic networks.

The following theorem implies that the asymptotic behaviors of the average \ac{cdi} in large stochastic networks assembles that of the \ac{cdi} in infinite lattice networks.

\begin{theorem}\label{thm:stochastic}
For $B<\infty$ and $R_\mathrm{max}= \Theta(B^{\frac{1}{k}})$ where $k>3$, the following convergence holds
\begin{equation}\label{sto_conv}
\mathbb{P}\Bigg\{\frac{\left|\frac{1}{N_\mathrm{a}}\!\left(\mathrm{tr}\left\{\RM{L}_{\RV{\theta},\mathrm{S}}\right\}\!-\!\mathrm{tr}\left\{\M{L}_{\RV{\theta},\mathrm{ L}}\right\}\right)\right|}{\frac{1}{N_\mathrm{a}}\mathrm{tr}\left\{\M{L}_{\RV{\theta},\mathrm{L}}\right\}-1} \!\neq\! o(1) \Bigg\}
\leq o\left(\bar{d}N_\mathrm{a}^{-3}\right)
\end{equation}
where $\M{L}_{\RV{\theta},\mathrm{L}}$ corresponds to a lattice network with the same $B$ as the stochastic network and maximum communication range $\sqrt{\lambda_\mathrm{s}}R_\mathrm{max}$, and is minimax matched \cite{minimax} with the stochastic one. Here we use the notation $\bar{d}$ to denote the expected number of neighboring agents of an agent in the interior area of the network, i.e., $\bar{d} = \pi R_\mathrm{max}^2$.
\begin{IEEEproof}
See Appendix \ref{sec:proof_stochastic}.
\end{IEEEproof}
\end{theorem}

Moreover, using the same technique as applied in the proof of Theorem \ref{thm:stochastic}, we have the following result on the asymptotic behavior of relative \ac{cdi}.
\begin{corollary}\label{coro:sto_rel}
For networks performing relative synchronization, under the same assumptions as in Theorem \ref{thm:stochastic}, the following convergence holds
\begin{equation}\label{sto_conv_rel}
\begin{aligned}
\mathbb{P}\Bigg\{\frac{\left|\frac{1}{N_\mathrm{a}}\tr{(\M{I}\!-\!\RM{P}_{\V{\theta},\mathrm{S}})^\dagger\!-\!(\M{I}\!-\!\M{P}_{\V{\theta},\mathrm{L}})^\dagger}\right|}{\frac{1}{N_\mathrm{ a}}\mathrm{tr}\left\{(\M{I}-\M{P}_{\V{\theta},\mathrm{L}})^\dagger\right\}-1}  \!\neq\! o(1) \Bigg\}\!\leq\! o\left(\bar{d}N_\mathrm{a}^{-3}\right).
\end{aligned}
\end{equation}
\begin{IEEEproof}
This corollary can be obtained straightforwardly using the techniques applied in Appendix \ref{sec:proof_stochastic}.
\end{IEEEproof}
\end{corollary}

\begin{remark}\label{rem:sto_asymptotic}
Theorem \ref{thm:stochastic} implies that the expected \ac{cdi} in a large stochastic network is of the same order as that in the minimax-matched lattice network. In other words, under the assumptions described in Theorem \ref{thm:stochastic}, Theorem \ref{thm:ici} can also be applied to sufficiently large stochastic networks in the ``convergence in probability'' sense and with high probability. Furthermore, from Corollary \ref{coro:sto_rel} we see that Proposition \ref{prop:rel_cdi} can also be applied to sufficiently large stochastic networks with high probability.
\end{remark}

\section{Numerical Results}\label{sec:numerical}
In this section, we illustrate and validate our previous analytical results using numerical examples.

\subsection{Scaling Laws}
All networks we consider in this subsection reside in a square area. The measurement variance $\sigma^2$ is chosen such that $2N\sigma^{-2}=1$, and the results are averaged over $1000$ realizations of the random network to approximate the expectation operation.

\begin{figure}[t]
    \centering
    \psfrag{z}[][]{\fitsize Expected \ac{rseb}}
    \includegraphics[width=.487\textwidth]{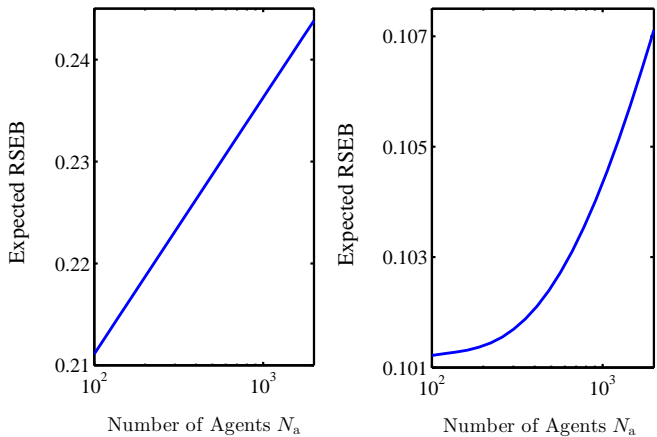}
    \caption{The expected \ac{rseb} $\mathbb{E}\big\{\bar{\rv{s}}(\V{\theta})\big\}$ as a function of the number of agents $N_\mathrm{a}$ in extended networks. Left: $R_{\max}\!=\!20$m. Right: $R_{\max}\!=\!25$m.}
    \label{fig:extended_rel}
\end{figure}

First we illustrate the scaling laws in extended networks. The intensity of agents is set as $\lambda_\mathrm{a} = 0.01$m$^{-2}$. Fig. \ref{fig:extended_rel} shows the expected \ac{rseb} $\mathbb{E}\big\{\bar{\rv{s}}(\V{\theta})\big\}$ as a function of $N_\mathrm{a}$. In the left plot the communication range $R_{\max}=20$m while in the right plot $R_{\max}=25$m. It can be seen that in both plots,  $\mathbb{E}\big\{\bar{\rv{s}}(\V{\theta})\big\}$ increases unboundedly as $N_\mathrm{a}$ grows, as shown in Proposition \ref{prop:extended_rel1}. Interestingly, as implied by Proposition \ref{prop:rel_cdi}, we also see that $\mathbb{E}\big\{\bar{\rv{s}}(\V{\theta})\big\}$ increases logarithmically as $N_\mathrm{a}$ grows when $N_\mathrm{a}$ is large.

\begin{figure}[t]
    \centering
    \psfrag{Expected ASEB}[][]{\fitsize Expected Average \ac{aseb}}
    \includegraphics[width=.45\textwidth]{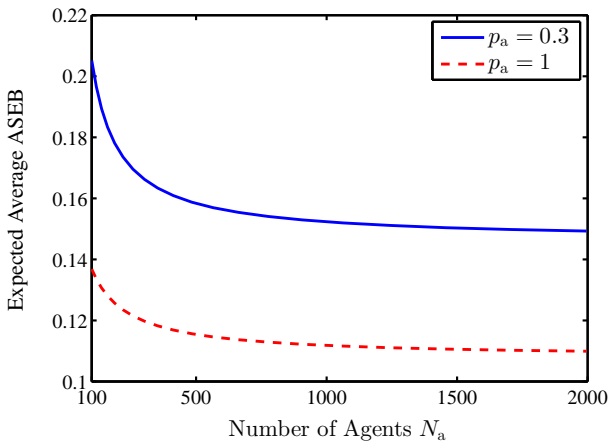}
    \caption{The expected average \ac{aseb} $\mathbb{E}\big\{N_\mathrm{a}^{-1}\sum_{i=1}^{N_\mathrm{ a}}\rv{s}(\rv{\theta}_i)\big\}$ as a function of the number of agents $N_\mathrm{a}$ in extended networks.}
    \label{fig:extended_abs}
\end{figure}

For networks performing absolute synchronization, we consider here the case that agents with \textit{a priori} information constitute a binomial point process with fixed intensity. The area of the network is $10000$m$^2$ and $R_{\max} = 20$m. Fig. \ref{fig:extended_abs} illustrates the expected average \ac{aseb} $\mathbb{E}\big\{N_\mathrm{a}^{-1}\sum_{i=1}^{N_\mathrm{ a}}\rv{s}(\rv{\theta}_i)\big\}$ as a function of $N_\mathrm{a}$, where agents have \textit{a priori} information with probability $p_\mathrm{a}=0.3$ and $p_\mathrm{a}=1$, respectively. In both cases, the expected average \ac{aseb} converges to a constant as $N_\mathrm{a}\rightarrow \infty$, as stated in Proposition \ref{prop:extended_abs2}. It is noteworthy that the expected average \ac{aseb} decreases before the convergence, which is a consequence of decreasing $n$-step returning probabilities caused by the extension of the network.

\begin{figure}[t]
    \centering
    \psfrag{z}[][]{\fitsize Expected \ac{rseb}}
    \includegraphics[width=.45\textwidth]{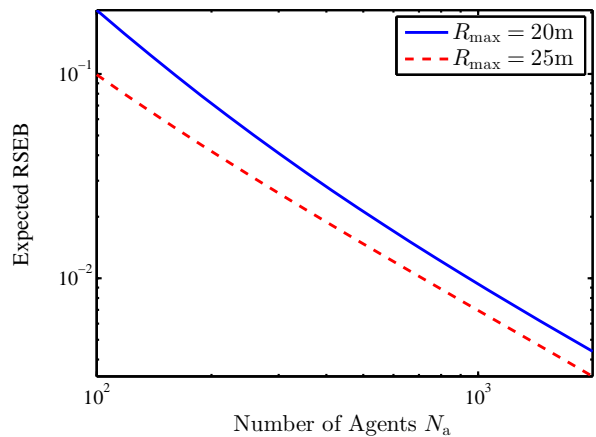}
    \caption{The expected \ac{rseb} $\mathbb{E}\big\{\bar{\rv{s}}(\V{\theta})\big\}$ as a function of the number of agents $N_\mathrm{a}$ in dense networks.}
    \label{fig:dense_rel}
\end{figure}

\begin{figure}[t]
    \centering
    \psfrag{Expected ASEB}[][]{\fitsize Expected Average \ac{aseb}}
    \includegraphics[width=.45\textwidth]{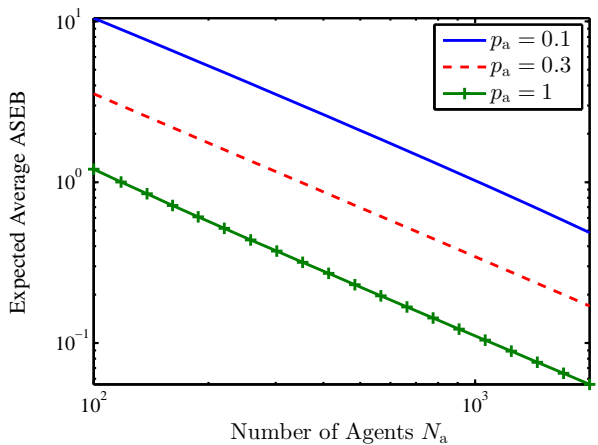}
    \caption{The expected average \ac{aseb} $\mathbb{E}\big\{N_\mathrm{a}^{-1}\sum_{i=1}^{N_\mathrm{ a}}\rv{s}(\rv{\theta}_i)\big\}$ as a function of the number of agents $N_\mathrm{a}$ in dense networks.}
    \label{fig:dense_abs}
\end{figure}

Next we investigate the scaling laws in dense networks. As can be seen in Fig. \ref{fig:dense_rel}, the expected \ac{rseb} $\mathbb{E}\big\{\bar{\rv{s}}(\V{\theta})\big\}$ scales as $\Theta(N_\mathrm{a}^{-1})$ for both $R_{\max}=20$m and $R_{\max}=25$m when $N_\mathrm{a}$ is large, which corroborates Proposition \ref{prop:dense_rel}. Similar behavior of the expected average \ac{aseb} is also seen in Fig. \ref{fig:dense_abs}, where it scales as $\Theta(N_\mathrm{a}^{-1})$ for various $p_\mathrm{a}$.

\subsection{Asymptotic Behavior of \ac{cdi}}
\begin{figure}[t]
    \centering
    \psfrag{Numerical eq. (34)}{\fitsize Numerical eq. \eqref{numerical}}
    \psfrag{Asymptotic eq. (80)}{\fitsize Asymptotic eq. \eqref{delta_ii_approx_pre}}
    \includegraphics[width=.45\textwidth]{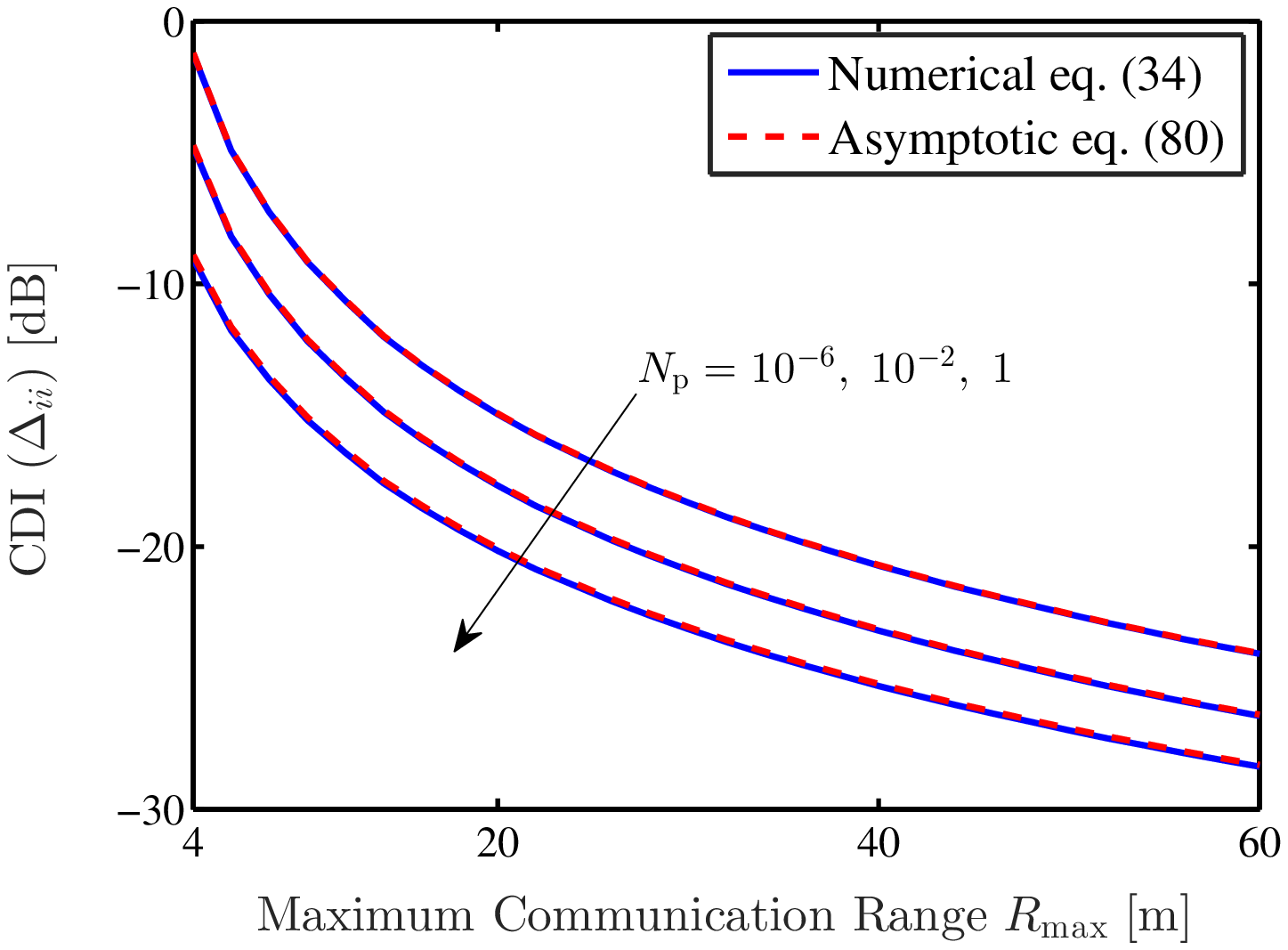}
    \caption{The \ac{cdi} $\Delta_{ii}$ as a function of $R_\mathrm{max}$ in infinite lattice networks, with different \np~$N_\mathrm{p}$.}
    \label{fig:delta_ii}
\end{figure}

In this subsection we demonstrate the results of analysis in Section \ref{sec:topology}. Without loss of generality, for all lattice networks, we consider a lattice size of $1$m$^2$. We first consider the behavior of \ac{cdi} as a function of $R_\mathrm{max}$ in infinite lattice networks. Figure \ref{fig:delta_ii} shows the numerical result with \np~$N_\mathrm{p} = 10^{-6}$, $10^{-2}$, and $1$. To emulate an ``infinite network'', we calculate the sum in \eqref{numerical} to its $n$-th term and calculate the residual using the Gaussian approximation \eqref{approximation_gaussian}. Parameter $n$ is chosen such that the relative approximation error of the $n$-th term given by
$$
\begin{aligned}
\frac{1}{p_{ij}^{(n)}}\bigg|&p_{ij}^{(n)}-\frac{1}{2\pi n\sigma_\mathrm{R}^2}\exp\Big\{-\frac{1}{n\sigma_\mathrm{R}^2}\|\V{p}_j-\V{p}_i\|^2\Big\}\bigg|
\end{aligned}
$$
is less than $10^{-3}$.

It can be observed that the asymptotic values of $\Delta_{ii}$ computed using \eqref{delta_ii_approx_pre} and the values obtained from numerical computation agree well even for small $R_{\max}$. The discrepancy between these two values diminishes as $R_{\max}$ increases. In addition, as \eqref{delta_ii_approx} states, when the information from the prior distribution is significant, $\Delta_{ii}$ is small since all agents can provide accurate timing information to their neighbors.

\begin{figure}[t]
    \centering
    \psfrag{Infinite Grid (Numerical)          }{\fitsize Inf. Lattice (Numerical) eq. \eqref{numerical}}
    \psfrag{Infinite Grid (Asymptotic)       }{\fitsize Inf. Lattice (Asymptotic) eq. \eqref{delta_ii_approx_pre}}
    \psfrag{Finite Grid (B=50)}{\fitsize Finite Lattice ($B\!=\!50$) eq. \eqref{delta_ij}}
    \psfrag{Finite Grid (B=100)}{\fitsize Finite Lattice ($B\!=\!100$) eq. \eqref{delta_ij}}
    \includegraphics[width=.45\textwidth]{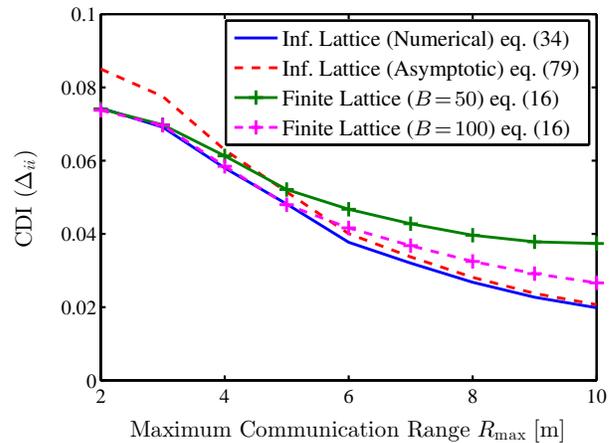}
    \caption{The \ac{cdi} $\Delta_{ii}$ as a function of $R_\mathrm{max}$ in finite lattice networks ($N_\mathrm{p}\!=\!5$ and $\lambda_\mathrm{s}\!=\!1$) compared with $\Delta_{ii}$ in infinite lattice networks ($N_\mathrm{p}\!=\!5$).}
    \label{fig:deltaii_fin_inf}
\end{figure}

Next we give an illustration of the analytical results on the average \ac{cdi} in Section \ref{ssec:fin_lattice}. We consider two different network sizes, namely $B=50$m and $B=100$m. For each case, the maximum communication range $R_\mathrm{max}$ varies from $2$m to $10$m, with fixed $N_\mathrm{p} = 5$ and node intensity $\lambda_\mathrm{s} = 1$. Figure \ref{fig:deltaii_fin_inf} shows the results. As a benchmark, we also plotted the $\Delta_{ii}$ of an infinite lattice network with the same $R_\mathrm{max}$ and $N_\mathrm{p}$, along with the asymptotic expressions presented in Section \ref{ssec:inf_lattice}.

As can be observed from Fig. \ref{fig:deltaii_fin_inf}, the gap between the \ac{cdi} in finite lattice networks and the \ac{cdi} in the infinite lattice network increases with $R_\mathrm{max}$ when $B$ is fixed. This can be understood as that the $n$-step transition probability of the Markov chain in finite lattice networks and that in infinite lattice networks are the same when $N \leq BR_\mathrm{max}^{-1}$. Therefore, the ``finite-infinite gap'' grows as $BR_\mathrm{max}^{-1}$ decreases. This can be further verified by noticing that the gap diminishes as $B$ increases when $R_\mathrm{max}$ is fixed. Furthermore, the \ac{cdi} in finite networks in Fig. \ref{fig:deltaii_fin_inf} can be seen as greater than the \ac{cdi} in the infinite lattice network, which corroborates Proposition \ref{prop:fin_coupling}.

\begin{figure}[t]
    \centering
    \psfrag{Finite Lattice (B=50) eq. (16)}{\fitsize Finite Lattice ($B\!=\!50$) eq. \eqref{delta_ij}}
    \psfrag{Finite Lattice (B=100) eq. (16)}{\fitsize Finite Lattice ($B\!=\!100$) eq. \eqref{delta_ij}}
    \psfrag{Stochastic (B=50)}{\fitsize Stochastic ($B\!=\!50$) eq. \eqref{stochastic_cdi}}
    \psfrag{Stochastic (B=100)}{\fitsize Stochastic ($B\!=\!100$) eq. \eqref{stochastic_cdi}}
    \includegraphics[width=.45\textwidth]{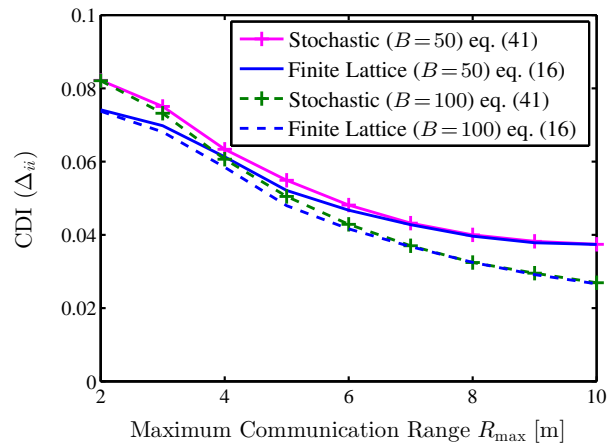}
    \caption{The expected \ac{cdi} as a function of $R_\mathrm{max}$ in stochastic networks ($N_\mathrm{p}=5$ and $\lambda_\mathrm{s} = 1$) compared with $\Delta_{ii}$ in finite lattice networks ($N_\mathrm{p}=5$).}
    \label{fig:delta_ii_sto_fin}
\end{figure}

Finally, we investigate the expected \ac{cdi} in stochastic networks discussed in Section \ref{ssec:inf_sto}. We use the same parameters as those used for investigating finite lattice networks, but the results for stochastic networks are averaged over 100 network snapshots. Simulation results are presented in Fig. \ref{fig:delta_ii_sto_fin}, where the \ac{cdi} of the corresponding finite lattice network is also plotted for comparison.

It can be seen that expected \ac{cdi} of stochastic networks converge to the average \ac{cdi} of the associated lattice network as $R_\mathrm{max}$ increases, as has been stated in Remark \ref{rem:sto_asymptotic}. If the abscissa is normalized by $B$, i.e., changed into $R_\mathrm{max}B^{-1}$, one can also find that the convergence rate from the expected \ac{cdi} of the stochastic network to the \ac{cdi} of the lattice network increases as $B$ increases.

\section{Conclusion}\label{sec:conclusions}
This paper develops a framework for the analysis of cooperative network synchronization. The general expression of the inverse \ac{fim} is proposed, based on which the concepts of \ac{cdi} and relative \ac{cdi} are introduced. We show that for absolute synchronization, reference nodes can be regarded as agents with infinite \textit{a priori} information, and agents with \textit{a priori} information can be regarded as that they can communicate with virtual reference nodes. To illustrate our framework, we provide random walk interpretations of both \ac{cdi} and relative \ac{cdi}. These interpretations are further exploited in the analysis on scaling laws for synchronization accuracy. We derive asymptotic expressions of \ac{cdi} in infinite lattice networks as well as relative \ac{cdi} in finite lattice networks, reflecting the relation between these quantities and network topology. Furthermore, we show that these results can also be applied to stochastic networks, and thus can be applied to real-world wireless networks. Our results reveal the relation between fundamental limits of synchronization accuracy and network-level system parameters, and will be useful in design and operation of wireless networks.

%---------------------------------------------------------------------------%
%                            acknowledgment                                 %
%---------------------------------------------------------------------------%
\section*{Acknowledgments}

The authors wish to thank A.~Conti, H.~Elsawy, Z.~Liu, and F.~Meyer for their helpful suggestions and careful reading of the manuscript.

%%%% APPENDICES BELOW %%%%

\appendices
\section{Proof of Proposition \ref{prop:structure}}\label{sec:proof_structure}
\begin{IEEEproof}
The structure of \ac{fim} can be obtain from \eqref{jlf2}. Note that the $(i,j)$-th entry of matrix $\M{J}_{\RV{\theta}}$ takes the following form
\begin{equation}\label{entries}
\left[\M{J}_{\RV{\theta}}\right]_{i,j} = -\mathbb{E}_{\RS{T},\RV{\theta}}\bigg\{\frac{\partial^2 \ln f_{\RS{T},\RV{\theta}}(\Set{T},\V{\theta})}{\partial \rv{\theta}_i \partial \rv{\theta}_j}\bigg\}.
\end{equation}
For off-diagonal entries we have
\begin{equation}\label{cross_inf}
\mathbb{E}_{\RS{T},\RV{\theta}}\bigg\{\frac{\partial^2 \ln f_{\RS{T},\RV{\theta}}(\Set{T},\V{\theta})}{\partial \rv{\theta}_i \partial \rv{\theta}_j}\bigg\} =
\left\{
  \begin{array}{ll}
    2N\sigma^{-2}, & \hbox{$j\in\Set{N}_i$} \\
    0, & \hbox{otherwise}
  \end{array}
\right.
\end{equation}
and for diagonal entries
\begin{equation}\label{self_inf}
\begin{aligned}
&\mathbb{E}_{\RS{T},\RV{\theta}}\bigg\{\frac{\partial^2 \ln f_{\RS{T},\RV{\theta}}(\Set{T},\V{\theta})}{\partial \rv{\theta}_i^2}\bigg\}\\
&\hspace{3mm}= -\xi_{\mathrm{P},i}-\sum_{k\in\Set{N}_i}\frac{2N}{\sigma^2}\\
&\hspace{3mm}= -\xi_{\mathrm{P},i}-\frac{2N}{\sigma^2}\left(\left|\Set{A}\cap\Set{N}_i\right| + \left|\Set{R}\cap\Set{N}_i\right|\right) \\
&\hspace{3mm}= -\xi_{\mathrm{P},i}-\frac{2N}{\sigma^2}(d_{{\rm A},i}+d_{{\rm R},i}.
\end{aligned}
\end{equation}
Substituting \eqref{cross_inf} and \eqref{self_inf} into \eqref{entries} yields \eqref{decompose}.
\end{IEEEproof}

\section{Proof of Theorem \ref{thm:decomposition}}\label{sec:proof_thm_main}
\begin{IEEEproof}
To obtain the inverse \ac{fim}, first note that the \ac{fim} $\M{J}_{\RV{\theta}}$ can be rewritten as below using \eqref{decompose}
\begin{equation}
\M{J}_{\RV{\theta}} = \frac{2N}{\sigma^2}\left(\M{D}_{\RV{\theta}}^\mathrm{C}+\M{D}_{\RV{\theta}}^\mathrm{R}\right)
+\M{\Xi}_{\RV{\theta}}^\mathrm{P}-\frac{2N}{\sigma^2}\M{A}_{\RV{\theta}}
\end{equation}
hence its inverse can be expressed as follows with the aid of the matrix inversion lemma
\begin{equation}\label{inversion_lemma}
\begin{aligned}
\M{J}_{\RV{\theta}}^{-1} =& \M{P}_{\RV{\theta}}(\M{I}-\M{P}_{\RV{\theta}})^{-1}\Big(\frac{2N}{\sigma^2}\big(\M{D}_{\RV{\theta}}^\mathrm{C}+\M{D}_{\RV{\theta}}^\mathrm{R}\big)
+\M{\Xi}_{\RV{\theta}}^\mathrm{P}\Big)^{-1} \\
&+\Big(\frac{2N}{\sigma^2}\big(\M{D}_{\RV{\theta}}^\mathrm{C}+\M{D}_{\RV{\theta}}^\mathrm{R}\big)
+\M{\Xi}_{\RV{\theta}}^\mathrm{P}\Big)^{-1}.
\end{aligned}
\end{equation}
By expanding the matrix $\left(\M{I}-\M{P}_{\RV{\theta}}\right)^{-1}$ as matrix power series
$$
\left(\M{I}-\M{P}_{\RV{\theta}}\right)^{-1} = \sum_{n=0}^{\infty}\M{P}_{\RV{\theta}}^n,
$$
we can rewrite \eqref{inversion_lemma} as
\begin{equation}
\M{J}_{\RV{\theta}}^{-1}=\bigg(\M{I}+\sum_{n=1}^{\infty}\M{P}_{\RV{\theta}}^n\bigg)\Big(\frac{2N}{\sigma^2}\big(\M{D}_{\RV{\theta}}^\mathrm{C}+\M{D}_{\RV{\theta}}^\mathrm{R}\big)
+\M{\Xi}_{\RV{\theta}}^\mathrm{P}\Big)^{-1}.
\end{equation}
Note that $\tilde{\M{D}}\triangleq\frac{2N}{\sigma^2}\left(\M{D}_{\RV{\theta}}^\mathrm{C}+\M{D}_{\RV{\theta}}^\mathrm{R}\right)
+\M{\Xi}_{\RV{\theta}}^\mathrm{P}$ is a diagonal matrix, therefore, the $(i,j)$-th entry in $\M{J}_{\RV{\theta}}^{-1}$ can be expressed as
\begin{equation}\label{pre_main}
\left[\M{J}_{\RV{\theta}}^{-1}\right]_{i,j} = \left\{
                                                 \begin{array}{ll}
                                                   (1+\Delta_{ii})\left[\tilde{\M{D}}\right]_{i,i}^{-1}, & \hbox{$i=j$} \\
                                                   \Delta_{ij}\left[\tilde{\M{D}}\right]_{j,j}^{-1}, & \hbox{$i\neq j$}
                                                 \end{array}
                                               \right.
\end{equation}
where $\Delta_{ij}$ is defined in \eqref{delta_ij}. Rearranging \eqref{pre_main} yields \eqref{main}.

Moreover, from the definition of $\M{P}_{\RV{\theta}}$ we see that all its entries are non-negative, as well as the entries of $\M{P}_{\RV{\theta}}^n$ for $n > 1$. Therefore we have $\Delta_{ij}\geq 0~\forall i,j$.
\end{IEEEproof}

\section{Proof of Proposition \ref{prop:equivalence}}\label{sec:proof_prop_equivalence}
\begin{IEEEproof}
In this proposition we investigate the behavior of agents with infinite \textit{a priori} information. Without loss of generality, consider a network with $N_\mathrm{a}$ agents whose corresponding \ac{fim} is given as \eqref{decompose}, and suppose that the agent $N_\mathrm{a}$ has infinite \textit{a priori} information such that $\xi_{\mathrm{P},N_\mathrm{a}}\rightarrow \infty$. Note that the matrix $\M{J}_{\RV{\theta}}$ can be partitioned as
\begin{equation}
\M{J}_{\RV{\theta}} = \left[
                         \begin{array}{cc}
                           \left[\M{J}_{\RV{\theta}}\right]_{\bar{N_\mathrm{a}}} & \left[\M{J}_{\RV{\theta}}\right]_{1:N_\mathrm{a}-1,N_\mathrm{a}}\\
                             \left[\M{J}_{\RV{\theta}}\right]_{1:N_\mathrm{a}-1,N_\mathrm{a}}^\mathrm{T} & [\M{J}_{\RV{\theta}}]_{N_\mathrm{a},N_\mathrm{a}} \\
                         \end{array}
                       \right].
\end{equation}
Thus the inverse \ac{fim} without the $N_\mathrm{a}$-th row and the $N_\mathrm{a}$-th column takes the following form
\begin{equation}
\left[\M{J}_{\RV{\theta}}^{-1}\right]_{\bar{N_\mathrm{a}}} =
\left([\M{J}_{\RV{\theta}}]_{\bar{N_\mathrm{a}}}\right)^{-1}-
\frac{\big\|\left[\M{J}_{\RV{\theta}}\right]_{1:N_\mathrm{a}-1,N_\mathrm{a}}\big\|^2}{[\M{J}_{\RV{\theta}}]_{N_\mathrm{a},N_\mathrm{a}}}
\end{equation}
which follows from the Schur complement lemma. According to \eqref{decompose}, the term $\left[\M{J}_{\RV{\theta}}\right]_{N_\mathrm{a},N_\mathrm{a}}$ can be expressed as \cite{coop_fundamental}
$$
\left[\M{J}_{\RV{\theta}}\right]_{N_\mathrm{a},N_\mathrm{a}} = \frac{2N}{\sigma^2}\left(d_{{\rm A},N_{\rm a}}+d_{{\rm R},N_{\rm a}}\right)+\xi_{\mathrm{ P},N_\mathrm{a}}
$$
and $\lim_{\xi_{\mathrm{P},N_\mathrm{a}}\rightarrow \infty}\left[\M{J}_{\RV{\theta}}\right]_{N_\mathrm{a},N_\mathrm{a}} = \infty$, hence
\begin{equation}
\lim_{\xi_{\mathrm{P},N_\mathrm{a}}\rightarrow \infty}\left[\M{J}_{\RV{\theta}}^{-1}\right]_{\bar{N_\mathrm{a}}} =
\left([\M{J}_{\RV{\theta}}]_{\bar{N_\mathrm{a}}}\right)^{-1}.
\end{equation}
\end{IEEEproof}

\section{Proof of Theorem \ref{thm:random_walk}}\label{sec:proof_random_walk}
\begin{IEEEproof}
To derive the random walk interpretation, we rewrite the extended \ac{fim} $\M{J}_{\tilde{\RV{\theta}}}$ as
\begin{equation}\label{tilde_j_theta}
\M{J}_{\tilde{\RV{\theta}}} = \M{D}_{\tilde{\RV{\theta}}}-\M{W}_{\tilde{\RV{\theta}}}
\end{equation}
where
\begin{equation}
\begin{aligned}
\left[\M{W}_{\tilde{\RV{\theta}}}\right]_{i,j} &= \left\{
                                      \begin{array}{ll}
                                        2N\sigma^{-2}, & \hbox{$j\in\Set{N}_i,~\{i,j\}\cap \Set{R}_\mathrm{v} = \emptyset$} \\
                                        \xi_{\mathrm{P},\min\{i,j\}}, & \hbox{$j\in\Set{N}_i,~\{i,j\}\cap \Set{R}_\mathrm{v} \neq \emptyset$} \\
                                        0, & \hbox{otherwise}
                                      \end{array}
                                    \right. \\
\left[\M{D}_{\tilde{\RV{\theta}}}\right]_{i,j} &=\left\{
                                                    \begin{array}{ll}
                                                      \sum_{k=1}^{2N_\mathrm{a}+N_\mathrm{r}}
\left[\M{W}_{\tilde{\RV{\theta}}}\right]_{i,j}, & \hbox{$i=j,~i\in\Set{A}$} \\
                                                     \xi_{\rm inf}, & \hbox{$i=j,~i\in\Set{R}\cup\Set{R}_\mathrm{v}$} \\
                                                      0, & \hbox{$i\neq j$}
                                                    \end{array}
                                                  \right.
\end{aligned}
\end{equation}
Using the same techniques applied in the derivation of Theorem \ref{thm:decomposition}, we can obtain
\begin{equation}
\left[\M{J}_{\tilde{\RV{\theta}}}^{-1}\right]_{i,j} = \left\{
                                                 \begin{array}{ll}
                                                   (1+\tilde{\Delta}_{ii})\left[\M{D}_{\tilde{\RV{\theta}}}\right]_{i,i}^{-1}, & \hbox{$i=j$} \\
                                                   \tilde{\Delta}_{ij}\left[\M{D}_{\tilde{\RV{\theta}}}\right]_{j,j}^{-1}, & \hbox{$i\neq j$}
                                                 \end{array}
                                               \right.
\end{equation}
where
\begin{subequations}
\begin{align}
\label{tilde_delta_ij}
\tilde{\Delta}_{ij} &\triangleq \sum_{n=1}^{\infty}\left[\tilde{\M{P}}_{\tilde{\RV{\theta}}}^n\right]_{i,j}, \\
\label{tilde_p_theta}
\tilde{\M{P}}_{\tilde{\RV{\theta}}} &\triangleq \lim_{\xi_{\rm inf}\rightarrow +\infty} \M{D}_{\tilde{\RV{\theta}}}^{-1}\M{W}_{\tilde{\RV{\theta}}}.
\end{align}
\end{subequations}
Thus from \eqref{tilde_j_theta} and \eqref{tilde_p_theta} we have
\begin{equation}
\left[\tilde{\M{P}}_{\tilde{\RV{\theta}}}\right]_{i,j} =  \left\{
                                                     \begin{array}{ll}
                                                       -\dfrac{\left[\M{J}_{\tilde{\RV{\theta}}}\right]_{i,j}}
{\left[\M{J}_{\tilde{\RV{\theta}}}\right]_{i,i}}, & \hbox{$i\neq j$;} \\
                                                       0, & \hbox{$i=j$.}
                                                     \end{array}
                                                   \right.
\end{equation}
But note that for $\Delta_{ij}$ defined as $\Delta_{ij}\triangleq \tilde{\Delta}_{ij},~i,j\in\Set{A}$, nothing will change if we replace the matrix $\tilde{\M{P}}_{\tilde{\RV{\theta}}}$ by $\M{P}_{\tilde{\RV{\theta}}}$ given by
\begin{equation}\label{tilde_delta_re}
\left[\M{P}_{\tilde{\RV{\theta}}}\right]_{i,j} =  \left\{
                                                     \begin{array}{ll}
                                                       -\dfrac{\left[\M{J}_{\tilde{\RV{\theta}}}\right]_{i,j}}
{\left[\M{J}_{\tilde{\RV{\theta}}}\right]_{i,i}}, & \hbox{$i\neq j$;} \\
                                                       0, & \hbox{$i=j,~i\in\Set{A}$;} \\
                                                       1, & \hbox{$i=j,~i\in\Set{R}\cup\Set{R}_{\rm v}$.}
                                                     \end{array}
                                                   \right.
\end{equation}
Matrix $\M{P}_{\tilde{\RV{\theta}}}$ is exactly the one-step transition matrix of the Markov chain mentioned in Theorem \ref{thm:random_walk}, implying the following result
\begin{equation}\label{N_step_transition}
\left[\M{P}_{\tilde{\RV{\theta}}}^n\right]_{i,j} = \mathbb{P}\{\rv{x}_n=j|\rv{x}_{0}=i\}.
\end{equation}
Note that $\mathbb{P}\{\rv{x}_k=j|\rv{x}_{k-1}=i\} = 0$ for $j\in\Set{A},i\in\Set{R}\cup\Set{R}_\mathrm{v}$ and $\mathbb{P}\{\rv{x}_k=i|\rv{x}_{k-1}=i\} = 1$ for $i\in\Set{R}\cup\Set{R}_\mathrm{v}$. Hence states in the set $\Set{R}\cup\Set{R}_\mathrm{v}$ are absorbing states of the Markov chain, and with \eqref{tilde_delta_ij}, \eqref{tilde_delta_re} and \eqref{N_step_transition} we obtain \eqref{sum_walk}.
\end{IEEEproof}

\section{Proof of Theorem \ref{thm:rel}}\label{sec:proof_rel}
\begin{IEEEproof}
We need an alternative form of the \ac{fim} $\M{J}_{\V{\theta}}$ to obtain the relative \ac{mse} bound. In the relative synchronization problem, we have $\M{P}_{\V{\theta}} = (\M{D}_{\V{\theta}}^\mathrm{C})^{-1}\M{A}_{\V{\theta}}$, and $\M{J}_{\V{\theta}}$ can thus be rewritten as
\begin{equation}\label{rel_jtheta}
\M{J}_{\V{\theta}} = \frac{\sigma^2}{2N}\M{D}_{\V{\theta}}^\mathrm{C}\left(\M{I}-\M{P}_{\V{\theta}}\right).
\end{equation}
Using \eqref{rel_jtheta}, the pseudo-inverse $\M{J}_{\V{\theta}}^\dagger$ can be expressed as
\begin{equation}\label{rel_pinv1}
\M{J}_{\V{\theta}}^\dagger = \frac{2N}{\sigma^2}\M{J}_{\V{\theta}}^\dagger\M{J}_{\V{\theta}}(\M{I}-\M{P}_{\V{\theta}})^{\rm g}\left(\M{D}_{\V{\theta}}^\mathrm{ C}\right)^{-1}\M{J}_{\V{\theta}}\M{J}_{\V{\theta}}^\dagger
\end{equation}
where the notation $\left[\cdot\right]^{\rm g}$ denotes a generalized inverse of its argument satisfying $\M{A}\M{A}^{\rm g}\M{A}=\M{A}$. Since the network is connected, $\M{J}_{\V{\theta}}$ only has a single eigenvector $N_\mathrm{a}^{-\frac{1}{2}}\V{1}_{N_\mathrm{a}}$ corresponding to eigenvalue $0$. Therefore we have $\M{J}_{\V{\theta}}^\dagger\M{J}_{\V{\theta}}=\M{J}_{\V{\theta}}\M{J}_{\V{\theta}}^\dagger =\M{C}$ where $\M{C}=\M{I}-N_\mathrm{a}^{-1}\V{1}_{N_\mathrm{a}}\V{1}_{N_\mathrm{a}}^\mathrm{T}$ is the centering matrix, and thus
\begin{equation}\label{rel_pinv}
\M{J}_{\V{\theta}}^\dagger=\frac{2N}{\sigma^2}\M{C}(\M{I}-\M{P}_{\V{\theta}})^{\rm g}\left(\M{D}_{\V{\theta}}^\mathrm{C}\right)^{-1}\M{C}.
\end{equation}

To find a suitable generalized inverse, consider the following matrix
$$
\M{Z}=\left(\M{I}-\M{P}_{\V{\theta}}+\M{\Pi}\right)^{-1}
$$
where $\M{\Pi}=\lim_{n\rightarrow \infty} \M{P}_{\V{\theta}}^n$. Note that $\M{Z}$ is a valid generalized inverse of $\M{I}-\M{P}_{\V{\theta}}$ since $\M{P}_{\V{\theta}}\M{\Pi}=\M{\Pi}\M{P}_{\V{\theta}}=\M{\Pi}$, and thus
\begin{equation}
\begin{aligned}
&(\M{I}-\M{P}_{\V{\theta}})\M{Z}(\M{I}-\M{P}_{\V{\theta}})\\
&\hspace{2mm}=(\M{I}-\M{P}_{\V{\theta}})\bigg(\M{I}+\sum_{n=1}^{\infty}(\M{P}_{\V{\theta}}^n-\M{\Pi})\bigg)(\M{I}-\M{P}_{\V{\theta}})\\
&\hspace{2mm}=\M{I}-\M{P}_{\V{\theta}}.
\end{aligned}
\end{equation}
According to the properties of the equilibrium distribution of Markov chains, matrix $\M{\Pi}$ can be expressed as $\M{\Pi}=\V{1}_{N_\mathrm{a}}\V{\pi}^\mathrm{T}$ where $\V{\pi}=[\pi_1~\pi_2~\dotsc~\pi_{N_\mathrm{a}}]^\mathrm{T}$ is the vector of equilibrium distribution. Thus we have
\begin{equation}\label{zd}
\Big[\M{Z}\left(\M{D}_{\V{\theta}}^\mathrm{C}\right)^{-1}\Big]_{i,j}=\frac{1}{d_{{\rm A},j}}\bigg\{1+\sum_{n=1}^\infty \left(\left[\M{P}_{\V{\theta}}^n\right]_{i,j}-\pi_j\right)\bigg\}.
\end{equation}
The trace of $\M{J}_{\V{\theta}}^\dagger$ can then be expressed as
$$
\begin{aligned}
\tr{\M{J}_{\V{\theta}}^\dagger}&=\frac{2N}{\sigma^2}\tr{\M{C}\M{Z}\left(\M{D}_{\V{\theta}}^\mathrm{C}\right)^{-1}} \\
&=\sum_{i=1}^{N_\mathrm{a}}\frac{1+\sum_{n=1}^\infty\left(\left[\M{P}_{\V{\theta}}^n\right]_{i,i}-\frac{1}{N_\mathrm{a}}\sum_{j=1}^{N_\mathrm{ a}}\left[\M{P}_{\V{\theta}}^n\right]_{j,i}\right)}{\frac{\sigma^2}{2N}d_{{\rm A},i}}
\end{aligned}
$$
yielding \eqref{rel_thm}.
\end{IEEEproof}

\section{Proof of Scaling Laws in Extended Networks}\label{sec:proof_scale_extended}
\subsection{Proof of Proposition \ref{prop:extended_rel1}}\label{sec:proof_extended_rel1}
\begin{IEEEproof}
In a connected network, each agent can communicate with at least one neighboring agent. Therefore we have
\begin{subequations}
\begin{align}
&\mathbb{E}\bigg\{\frac{1}{N_\mathrm{a}}\mathrm{tr}\big\{\RM{J}_{\V{\theta}}^\dagger\big\}\bigg\} \\
&\hspace{2mm}> \frac{\sigma^{2}}{2N}\mathbb{E}\bigg\{\frac{1}{N_\mathrm{a}}\sum_{i=1}^{N_\mathrm{a}}\tilde{\rv{\Delta}}_{ii}\bigg\}\\
&\hspace{2mm}\!=\frac{\sigma^{2}}{2N}\mathbb{E}\bigg\{\frac{1}{N_\mathrm{a}}\!\sum_{i=1}^{N_\mathrm{a}}\!\sum_{n=1}^\infty\!\bigg\{\left[\RM{P}_{\V{\theta}}^n\right]_{i,i}\!-\!\frac{1}{N_\mathrm{ a}}\!\sum_{j=1}^{N_\mathrm{a}}\!\left[\RM{P}_{\V{\theta}}^n\right]_{j,i}\!\bigg\}\bigg\}\label{fromrel}
\end{align}
\end{subequations}
where \eqref{fromrel} follows from Theorem \ref{thm:rel}. Rearranging the terms, the expectation can be rewritten as
\begin{subequations}\label{exp_rel_cdi}
\begin{align}
&\mathbb{E}\bigg\{\frac{1}{N_\mathrm{a}}\sum_{i=1}^{N_{\rm a}}\sum_{n=1}^\infty\bigg\{\left[\RM{P}_{\V{\theta}}^{n}\right]_{i,i}-\frac{1}{N_\mathrm{ a}}\sum_{j=1}^{N_{\rm a}}\left[\RM{P}_{\V{\theta}}^{n}\right]_{j,i}\bigg\}\bigg\} \\
&\hspace{2mm}=\!\sum_{n=1}^\infty\mathbb{E}\bigg\{\!\frac{1}{N_\mathrm{a}}\!\sum_{i=1}^{N_\mathrm{a}}\left[\RM{P}_{\V{\theta}}^n\right]_{i,i}\!-\!\frac{1}{N_\mathrm{a}^2}\!\sum_{j=1}^{N_\mathrm{ a}}\!\sum_{i=1}^{N_\mathrm{a}}\!\left[\RM{P}_{\V{\theta}}^n\right]_{j,i}\!\bigg\}\\
&\hspace{2mm}=\sum_{n=1}^\infty\bigg\{\mathbb{E}\bigg\{\frac{1}{N_\mathrm{a}}\tr{\RM{P}_{\V{\theta}}^n}\bigg\}-\frac{1}{N_\mathrm{a}}\bigg\}.\label{diverge_rel}
\end{align}
\end{subequations}
Note that as $N_{\rm a}\rightarrow \infty$, for any given $n$, the expected average $n$-step return probability $\E{N_\mathrm{a}^{-1}\tr{\RM{P}_{\V{\theta}}^n}}$ in the extended networks tends to its counterpart in an infinitely large network, which is a constant with respect to $N_{\rm a}$. Hence as $N_{\rm a}\rightarrow \infty$,
\begin{equation}
\sum_{n=1}^\infty\E{\frac{1}{N_{\rm a}}\tr{\RM{P}_{\V{\theta}}^n}} \sim \sum_{n=1}^\infty\bigg\{\E{\frac{1}{N_{\rm a}}\tr{\RM{P}_{\V{\theta}}^n}}-\frac{1}{N_{\rm a}}\bigg\}.
\end{equation}

Moreover, we have $\sum_{n=1}^\infty\E{N_\mathrm{a}^{-1}\tr{\RM{P}_{\V{\theta}}^n}}=\infty$ since $\RM{J}_{\V{\theta}}$ is not invertible. Therefore, as $N_\mathrm{a}\rightarrow\infty$, the sum of the series in \eqref{diverge_rel} tends to infinity, and thus the proof is completed.
\end{IEEEproof}

\subsection{Proof of Proposition \ref{prop:extended_abs1}}\label{sec:proof_extended_abs1}
\begin{IEEEproof}
Using Theorem \ref{thm:decomposition}, the following can be obtained
\begin{subequations}
\begin{align}
&\E{\rv{s}(\rv{\theta}_i)} \\
&\hspace{2mm}\geq \mathbb{E}\bigg\{\frac{1+\rv{\Delta}_{ii}}{\frac{2N}{\sigma^{2}}\rdeg{i}+\xi_{\max}}\bigg\}\\
&\hspace{2mm}\geq \frac{1}{\frac{2N}{\sigma^{2}}\E{\!\rdeg{i}\!}\!+\!\xi_{\max}}\!
+\!\mathbb{E}\bigg\{\!\frac{\rv{\Delta}_{ii}}{\frac{2N}{\sigma^{2}}\rdeg{i}\!+\!\xi_{\max}}\!\bigg\} \\
&\hspace{2mm}=\Omega(1)\label{lastline}
\end{align}
\end{subequations}
where \eqref{lastline} follows from the fact that $\rv{\Delta}_{ii}\geq 0$.
\end{IEEEproof}

\subsection{Proof of Proposition \ref{prop:extended_abs2}}\label{sec:proof_extended_abs2}
\begin{IEEEproof}
In the light of Proposition \ref{prop:extended_abs1}, it suffices to proof that $\lim_{N_\mathrm{a}\rightarrow \infty}\E{\rv{s}(\rv{\theta}_i)}<\infty$. Since the network is connected, the proof reduces to showing that $\lim_{N_\mathrm{a}\rightarrow \infty}\E{\rv{\Delta}_{ii}}<\infty$.

Now recall that the agents with \textit{a priori} information constitute a binomial point process $\rv{\Phi}(\cdot)$ with intensity $\lambda_\mathrm{ap}$ fixed as $N_\mathrm{a}\rightarrow\infty$. With this assumption, the probability that there is no agent with \textit{a priori} information in a given region $\Set{R}$ can be calculated as
\begin{equation}
\Prob{\rv{\Phi}(\Set{R})=0}=\frac{(|\Set{R}_\mathrm{net}|-|\Set{R}|)^{N_\mathrm{a}}}{|\Set{R}_\mathrm{net}|^{N_\mathrm{a}}}
\end{equation}
which tends to zero as $N_\mathrm{a}\rightarrow \infty$. Therefore as $N_\mathrm{a}\rightarrow \infty$, we have that for any agent $i$, $\exists n<\infty,~\left[\RM{P}_{\RV{\theta}}^n\right]_{i,j}>0$ with probability approaching $1$. According to the discussion in Remark \ref{abs_syncable}, this implies that the \ac{cdi} is finite with probability approaching $1$, and thus the expected \ac{cdi} $\E{\rv{\Delta}_{ii}}$ is guaranteed to be finite. Hence the proof is completed.
\end{IEEEproof}

\subsection{Proof of Proposition \ref{prop:dense_rel}}\label{sec:proof_dense_rel}
\begin{IEEEproof}
From \eqref{exp_rel_cdi} we see that the average expected relative \ac{cdi} is given by
\begin{equation}
\begin{aligned}
\mathbb{E}\bigg\{\frac{1}{N_\mathrm{a}}\sum_{i=1}^{N_\mathrm{a}}\tilde{\rv{\Delta}}_{ii}\bigg\}&\!=\!\sum_{n=1}^\infty\bigg\{\mathbb{E}\bigg\{\frac{1}{N_\mathrm{ a}}\tr{\RM{P}_{\V{\theta}}^n}\bigg\}\!-\!\frac{1}{N_\mathrm{a}}\bigg\}\\
&\!=\!\sum_{n=1}^\infty\mathbb{E}\bigg\{\!\frac{1}{N_\mathrm{a}}\left(\tr{\RM{P}_{\V{\theta}}^n}\!-\!\tr{\RM{P}_{\V{\theta}}^\infty}\right)\!\bigg\}.
\end{aligned}
\end{equation}
It is obvious that $\tr{\RM{P}_{\V{\theta}}}=0$. For any given $n\geq 2$,
\begin{equation}
\begin{aligned}
&\frac{1}{N_\mathrm{a}}\tr{\RM{P}_{\V{\theta}}^n}-\frac{1}{N_\mathrm{a}}\tr{\RM{P}_{\V{\theta}}^\infty} =\frac{1}{N_\mathrm{a}}\sum_{i=2}^{N_\mathrm{a}}\lambda_i^n(\RM{P}_{\V{\theta}})\\
&\hspace{2mm}\leq \lambda_\mathrm{slem}^{n-2}\left|\frac{1}{N_\mathrm{a}}\tr{\RM{P}_{\V{\theta}}^2}-\frac{1}{N_\mathrm{a}}\tr{\RM{P}_{\V{\theta}}^\infty}\right|
\end{aligned}
\end{equation}
where $\lambda_i(\M{A})$ denotes the $i$-th largest eigenvalue of $\M{A}$ and $\lambda_\mathrm{slem}=\max\{\lambda_2(\RM{P}_{\V{\theta}}),|\lambda_{n-1}(\RM{P}_{\V{\theta}})|\}$ is the second largest eigenvalue modulus (SLEM) of $\RM{P}_{\V{\theta}}$. Thus
\begin{equation}\label{rel_dense_cdi}
\begin{aligned}
\frac{1}{N_\mathrm{a}}\sum_{i=1}^{N_\mathrm{a}}\!\tilde{\rv{\Delta}}_{ii}&\!\leq \!\frac{1}{1-\lambda_\mathrm{slem}}\!\cdot\!\frac{1}{N_\mathrm{a}}\tr{\RM{P}_{\V{\theta}}^2}\\
&\!=\!\frac{1}{1-\lambda_\mathrm{slem}}\!\cdot\!\frac{1}{N_\mathrm{a}}\sum_{i=1}^{N_\mathrm{a}}\!\sum_{j\in\Set{N}_i}\!\frac{1}{\rdeg{i}\rdeg{j}}.
\end{aligned}
\end{equation}
According to \cite{boyd2005mixing}, $1-\lambda_\mathrm{slem}$ has a lower bound independent of $N_\mathrm{a}$ in dense networks. It can then be observed from \eqref{rel_dense_cdi} that $\E{N_\mathrm{a}^{-1}\sum_{i=1}^{N_\mathrm{a}}\tilde{\rv{\Delta}}_{ii}}=O(N_\mathrm{a}^{-1})$ since $(\rdeg{i})^{-1}$ scales as $\Theta(N_\mathrm{a}^{-1})$ with probability approaching $1$ as $N_\mathrm{a}\rightarrow \infty$. Hence $\E{N_\mathrm{a}^{-1}\tr{\RM{J}_{\V{\theta}}^\dagger}}=\Theta(N_\mathrm{a}^{-1})$ can be obtained from \eqref{rel_thm}.
\end{IEEEproof}

\subsection{Proof of Proposition \ref{prop:dense_abs}}\label{sec:proof_dense_abs}
\begin{IEEEproof}
Using similar arguments as applied in Proposition \ref{prop:extended_abs2}, it can be shown that under aforementioned assumptions, the \ac{cdi} $\rv{\Delta}_{ii}$ is finite for any agent $i$ with probability approaching $1$ as $N_\mathrm{a}\rightarrow \infty$. Thus $\E{N_\mathrm{a}^{-1}\sum_{i=1}^{N_\mathrm{a}}\rv{s}(\rv{\theta}_i)}$ scales as $\Theta(N_\mathrm{a}^{-1})$ since
\begin{equation}
\mathbb{E}\bigg\{\frac{1}{N_\mathrm{a}}\sum_{i=1}^{N_\mathrm{a}}\frac{1}{\rdeg{i}}\bigg\}=\Theta(N_\mathrm{a}^{-1})
\end{equation}
holds in dense networks.
\end{IEEEproof}

\section{Proof of Theorem \ref{thm:ici}}\label{sec:proof_ici}
\begin{IEEEproof}
To obtain an asymptotic expression of $\Delta_{ii}$, we start from expressing it as
\begin{equation}
\begin{aligned}
\Delta_{ii} &\!=\! \sum_{n=2}^{\infty} \mathbb{P}\{\RV{x}_n=\V{x}_0|\RV{x}_0=\V{x}_0\} \Bigg(\frac{\bar{d}}{\bar{d}+\frac{\sigma^2\xi_\mathrm{P}}{2N}}\Bigg)^n \\
&\!=\! \frac{1}{2\pi \sigma_\mathrm{R}^2}\sum_{n=2}^{\infty} \frac{1+E_1(n,\|\V{x}-\V{x}_0\|)}{n}\Bigg(\frac{\bar{d}}{\bar{d}+\frac{\sigma^2\xi_\mathrm{P}}{2N}}\Bigg)^n \\
&\!=\! \frac{1}{2\pi \sigma_\mathrm{R}^2}\sum_{n=2}^{\infty}\gamma_{ii}(n,\bar{d})+\epsilon_{ii}(n,\bar{d})
\end{aligned}
\end{equation}
with
\begin{equation}
\begin{aligned}
\gamma_{ii}(n,\bar{d})&\triangleq\frac{1}{n}\Bigg(\frac{\bar{d}}{\bar{d}+\frac{\sigma^2\xi_\mathrm{P}}{2N}}\Bigg)^n, \\
\epsilon_{ii}(n,\bar{d})&\triangleq \frac{E_1(n,\|\V{x}-\V{x}_0\|)}{n}\Bigg(\frac{\bar{d}}{\bar{d}+\frac{\sigma^2\xi_\mathrm{P}}{2N}}\Bigg)^n .
\end{aligned}
\end{equation}
We next show that
\begin{equation}\label{order_equivalence}
\lim_{\bar{d}\rightarrow \infty} \frac{\sum_{n=2}^{\infty}\epsilon_{ii}(n,\bar{d})}
{\sum_{n=2}^{\infty}\gamma_{ii}(n,\bar{d})} = 0.
\end{equation}
Since $\lim_{n\rightarrow \infty} E_1(n,\|\V{x}-\V{x}_0\|) =0$, we have
$$
\lim_{n\rightarrow \infty} \frac{\epsilon_{ii}(n,\bar{d})}{\gamma_{ii}(n,\bar{d})} = 0
$$
independent of $\bar{d}$. Therefore, we can arbitrarily choose a sufficiently large number $M$, such that
$$
\frac{\epsilon_{ii}(n,\bar{d})}{\gamma_{ii}(n,\bar{d})} \leq k~\forall n>M.
$$
The left hand side in \eqref{order_equivalence} can now be expressed as
\begin{subequations}
\begin{align}
&\lim_{\bar{d}\rightarrow \infty} \frac{\sum_{n=2}^{\infty}\epsilon_{ii}(n,\bar{d})}
{\sum_{n=2}^{\infty}\gamma_{ii}(n,\bar{d})} \\
&\hspace{3mm}= \lim_{\bar{d}\rightarrow \infty} \frac{\sum_{n=2}^{M}\epsilon_{ii}(n,\bar{d})+\sum_{n=M+1}^{\infty}\epsilon_{ii}(n,\bar{d})}
{\sum_{n=2}^{\infty}\gamma_{ii}(n,\bar{d})} \\
&\hspace{3mm}\label{decomp_series}\leq k
\end{align}
\end{subequations}
where \eqref{decomp_series} follows from the fact that the sum of finite terms of $e_{ii}(n,\bar{d})$ is finite. Since $M$ is arbitrary, the limit is actually zero, hence the result below follows
\begin{equation}\label{delta_ii_approx_pre}
\begin{aligned}
\Delta_{ii} &\sim \frac{1}{2\pi\sigma_\mathrm{R}^2}\sum_{n=2}^{\infty}\gamma_{ii}(n,\bar{d}) \\
&=\frac{1}{2\pi\sigma_\mathrm{R}^2}\bigg\{\ln\bigg(1+\frac{2N}{\sigma^2\xi_\mathrm{P}}\cdot\bar{d}\bigg)-
\frac{\bar{d}}{\bar{d}+\frac{\sigma^2\xi_\mathrm{P}}{2N}}\bigg\}.
\end{aligned}
\end{equation}
With some tedious but straightforward calculations, we have
\begin{equation}\label{sigma}
\sigma_\mathrm{R}^2 \sim \frac{\bar{d}}{4\pi}.
\end{equation}
Substituting \eqref{sigma} into \eqref{delta_ii_approx_pre}, we obtain \eqref{delta_ii_approx}.
\end{IEEEproof}

\section{Proof of Proposition \ref{prop:fin_coupling}}\label{sec:proof:fin}
\begin{IEEEproof}
In this section, we show that the \ac{cdi} in finite lattice networks is no less than that in infinite lattice networks under certain conditions, by means of mathematical induction. Denote by $f_{i\Set{A}}^{(n)}(j)$ the probability of the event ``starting from agent $i$, the random walker arrives at agent $j$ at the $n$-th step, which is the first time it arrives in set $\Set{A}$''. To prove $\Delta_{ii,\mathrm{L}} \geq \Delta_{ii,\mathrm{IL}}~\forall i\in\Set{I}$, it suffices to show that
$$
p_{ii,\mathrm{L}}^{(n)}\geq p_{ii,\mathrm{IL}}^{(n)}~\forall n\in\mathbb{Z}_+,~i\in\Set{I}.
$$
Denote the subscript set $\{\mathrm{L, IL}\}$ as $\Set{S}$, we have
$$
\begin{aligned}
p_{ii,z}^{(n)} =& \mathbb{P}_z\{\rv{x}_n=i|\rv{x}_0=i,x_j\notin \Set{E}~\forall j\leq n\} \\
&+\sum_{j\in\Set{E}}\sum_{k=1}^{n-1}f_{i\Set{E},z}^{(k)}(j)p_{ji,z}^{(n-k)},~z\in\Set{S}.
\end{aligned}
$$
Note that $\mathbb{P}_z\{\rv{x}_n=i|\rv{x}_0=i,x_j\notin \Set{E}~\forall j\leq n\}$ and $f_{i\Set{E},z}^{(k)}(j)$ are identical for $z=\mathrm{L}$ and $z=\mathrm{IL}$, and thus the proof can be reduced to showing that
$$
p_{ji,\mathrm{L}}^{(n)}\geq p_{ji,\mathrm{IL}}^{(n)}~\forall i\in\Set{I},~j\in\Set{E},~n\in\mathbb{Z}_+.
$$
For given $n\geq 2$, we have
\begin{equation}
\begin{aligned}
p_{ji,\mathrm{L}}^{(n)} &\!=\! \sum_{k\in\Set{N}_{j,\mathrm{L}}}p_{ji,\mathrm{L}}^{(1)}p_{ki,\mathrm{L}}^{(n-1)} \\
&\!=\!\frac{1}{\left|\Set{N}_{j,\mathrm{L}}\right|}\sum_{k\in\Set{N}_{j,\mathrm{L}}}p_{ki,\mathrm{L}}^{(n-1)}~\forall i\in\Set{I},~j\in\Set{E}
\end{aligned}
\end{equation}
\begin{equation}\label{fln_1}
\begin{aligned}
p_{ji,\mathrm{Il}}^{(n)} &\!=\! \sum_{k\in\Set{N}_{j,\mathrm{IL}}}p_{jk,\mathrm{IL}}^{(1)}p_{ki,\mathrm{IL}}^{(n-1)} \\
&\!=\!\frac{1}{\left|\Set{N}_{j,\mathrm{IL}}\right|}\Big(\sum_{k\in\Set{N}_{j,\mathrm{L}}}p_{ki,\mathrm{IL}}^{(n-1)}+\sum_{k\in\Set{N}_{j,\mathrm{IL}}\backslash\Set{N}_{j,\mathrm{ L}}}p_{ki,\mathrm{IL}}^{(n-1)}\Big) \\
&\!\le\! \sum_{k\in\Set{N}_{j,\mathrm{L}}}\frac{p_{ki,\mathrm{IL}}^{(n-1)}}{\left|\Set{N}_{j,\mathrm{L}}\right|}~\forall i\in\Set{I},~j\in\Set{E}.
\end{aligned}
\end{equation}
The last line of \eqref{fln_1} follows from assumption \ref{as2}. Thus for any given $n\geq 2$, to prove $p_{ji,\mathrm{L}}^{(n)}\geq p_{ji,\mathrm{IL}}^{(n)}~\forall i\in\Set{I},~j\in\Set{E}$, it suffices to prove that
\begin{equation}\label{fln_2}
p_{ki,\mathrm{L}}^{(n-1)}\geq p_{ki,\mathrm{IL}}^{(n-1)}~\forall i\in\Set{I},~k\in\Set{G}_\mathrm{L}.
\end{equation}
Note that $p_{ki,\mathrm{L}}^{(1)} = p_{ki,\mathrm{IL}}^{(1)}~\forall k\in\Set{I},~i\in\Set{I}$, and hence to prove \eqref{fln_2} for any given $n\geq 3$, it suffices to show that
$$
p_{ki,\mathrm{L}}^{(n-2)}\geq p_{ki,\mathrm{IL}}^{(n-2)}~\forall i\in\Set{I},~k\in\Set{E}.
$$
As we repeat the recursion, for any given pair of agent $(k,i),~i\in\Set{I}$ in $\Set{G}_\mathrm{IL}$, the random walker will reach a point where all paths of $(n-m),~m\geq m^*$ steps must resides in $\Set{G}_\mathrm{L}$, according to assumption \ref{as1}. Therefore we have
$$
P_{ki,\mathrm{L}}^{(n-m)}=p_{ki,\mathrm{IL}}^{(n-m)},~m\geq m^*.
$$
The proof is thus completed.
\end{IEEEproof}

\section{Proof of Proposition \ref{prop:rel_cdi}}\label{sec:proof_rel_cdi}
\begin{IEEEproof}
To investigate the asymptotic behavior of relative \ac{cdi}, first we revisit \eqref{diverge_rel} in the derivation of scaling laws, and obtain
$$
\frac{1}{N_\mathrm{a}}\sum_{i=1}^{N_\mathrm{a}}\tilde{\Delta}_{ii} = \sum_{n=1}^\infty\bigg\{\frac{1}{N_\mathrm{a}}\sum_{i=1}^{N_\mathrm{a}}p_{ii,\mathrm{L}}^{(n)}-\frac{1}{N_\mathrm{ a}}\bigg\}.
$$
According to Proposition \ref{prop:fin_coupling}, we see that for any agent $i$ in the interior area, $p_{ii,\mathrm{L}}^{(n)}\geq p_{ii,\mathrm{IL}}^{(n)}~\forall n\in\mathbb{Z}_+$. Therefore we have
$$
\frac{1}{N_\mathrm{a}}\sum_{i=1}^{N_\mathrm{a}}p_{ii,\mathrm{L}}^{(n)} \ge\frac{1}{N_\mathrm{a}}\sum_{i=1}^{N_\mathrm{a}} p_{ii,\mathrm{IL}}^{(n)},~\forall n\in\mathbb{Z}_+.
$$
Hence
$$
\frac{1}{N_\mathrm{a}}\sum_{i=1}^\infty\tilde{\Delta}_{ii} \geq \sum_{n=1}^M\bigg\{\frac{1}{N_\mathrm{a}}\sum_{i=1}^{N_\mathrm{a}}p_{ii,\mathrm{IL}}^{(n)}-\frac{1}{N_\mathrm{a}}\bigg\} + o(1)
$$
where $M$ is chosen such that $p_{ii,\mathrm{IL}}^{(n)}\leq N_\mathrm{a}^{-1}~\forall n\geq M$. From Theorem \ref{thm:ici} we see that $p_{ii,\mathrm{IL}}^{(n)}=\Theta(n^{-1})+o(n^{-1})$, and thus
$$
\begin{aligned}
\sum_{n=1}^M\bigg\{\frac{1}{N_\mathrm{a}}\sum_{i=1}^{N_\mathrm{a}}p_{ii,\mathrm{IL}}^{(n)}-\frac{1}{N_\mathrm{a}}\bigg\} = \frac{1}{N_\mathrm{a}}\sum_{i=1}^{N_\mathrm{a}}\sum_{n=1}^M \frac{c_{in}}{n\bar{d}} + \mathrm{const}
\end{aligned}
$$
where $c_{in}$'s are constants. Note that $M=\Theta(N_\mathrm{a})$, we can now obtain \eqref{rel_cdi1} since $\sum_{n=1}^M n^{-1}=\Theta(\ln M)$.
\end{IEEEproof}

\setcounter{tempEqCounter}{\value{equation}}
\setcounter{equation}{91}
\begin{figure*}[t]
\begin{subequations}\label{converge_sto_sto}
\begin{align}
\left\|\RM{L}_{\RV{\theta},{\rm S}}^{-1} - \tilde{\RM{L}}_{\RV{\theta},{\rm S}}^{-1}\right\|_{\rm HS}&=\left\{\sum_{i\in\mathcal{A}} \frac{\tilde{\rv{d}}_i(\bm{p}_i)}{N_{\rm a}} \left(\frac{N_{\rm p}\left(\tilde{\rv{d}}_i(\bm{p}_i)-\bar{d}+\rv{e}_i\right)}{\left(\tilde{\rv{d}}_i(\bm{p}_i)+\rv{e}_i\right)\left(\bar{d}+N_{\rm p}\right)
\left(\tilde{\rv{d}}_i(\bm{p}_i)+\rv{e}_i+N_{\rm p}\right)}\right)^2\right\}^{\frac{1}{2}}\\
&\le \frac{1}{N_{\rm a}} \left\{\sum_{\scriptstyle i\in\mathcal{A} \atop \scriptstyle \|\bm{p}_i\|_{\infty}\le B-R_{\rm max}} \frac{N_{\rm p}\rv{e}_i\sqrt{\tilde{\rv{d}}_i(\bm{p}_i)}}{\left(\tilde{\rv{d}}_i(\bm{p}_i)+\rv{e}_i\right)\left(\bar{d}+N_{\rm p}\right)
\left(\tilde{\rv{d}}_i(\bm{p}_i)+\rv{e}_i+N_{\rm p}\right)}\right. \\
& \left.\hspace{9mm} + \sum_{\scriptstyle i\in\mathcal{A} \atop \scriptstyle \|\bm{p}_i\|_{\infty}> B-R_{\rm max}} \frac{\sqrt{\tilde{\rv{d}}_i(\bm{p}_i)}N_{\rm p}\left(\frac{3}{4}\bar{d}+\rv{e}_i\right)}{\left(\tilde{\rv{d}}_i(\bm{p}_i)+\rv{e}_i\right)\left(\bar{d}+N_{\rm p}\right)
\left(\tilde{\rv{d}}_i(\bm{p}_i)+\rv{e}_i+N_{\rm p}\right)}\right\} \\
&\le \frac{1}{N_{\rm a}}\left(t_1^2N_{{\rm int}}\bar{d}^{-2}\sqrt{\log \bar{d}}+t_2^2\left(N_{\rm a}-\rv{n}_{\rm int}\right)\bar{d}^{-\frac{3}{2}}\sqrt{\log \bar{d}}\right) \label{conv_wp}
\end{align}
\end{subequations}
\hrulefill
\end{figure*}
\setcounter{equation}{\value{tempEqCounter}}

\section{Proof of Theorem \ref{thm:stochastic}}\label{sec:proof_stochastic}
\begin{IEEEproof}
To prove the convergence of stochastic networks to lattice networks, we need the following lemma.
\begin{lemma}\label{lemma:concentrate}
For $R_\mathrm{max} = O(2\ln^{\frac{3}{4}}B)$ and positive $t$, there exists positive constant $c$ such that the following inequality holds
\begin{equation}\label{concentrate}
\begin{aligned}
\mathbb{P}&\left\{\left\|\mu(\RM{T}_{\RV{\theta},\mathrm{S}})-\mu(\M{T}_{\RV{\theta},\mathrm{L}})\right\|_\mathrm{WS}
> o\left(N_\mathrm{a}^{-\frac{1}{4}}\sqrt{R_{\max}}\right)\right\} \\
&\leq o\left(N_\mathrm{a}^{\frac{5}{4}}\sqrt{R_{\max}}\exp\left\{-N_\mathrm{a}R_{\max}^2\right\}\right)
\end{aligned}
\end{equation}
where $\mu(\M{A})$ is the spectral measure of $\M{A}\in\mathbb{R}^{n\times n}$ defined as
$$
\mu(\M{A}) \triangleq \frac{1}{n}\left|\{\lambda\in\mathrm{Spec}(\M{A}):\lambda\leq x\}\right|
$$
and $\|\cdot\|_\mathrm{WS}$ is the Wasserstein distance given by
$$
\|\mu-\nu\|_\mathrm{WS} \triangleq \sup_{f \in \Set{L}} \Big|\int f\mathrm{d}\mu - \int f \mathrm{d} \nu\Big|
$$
with $\Set{L}$ denotes the set of all Lipschitz continuous functions. The matrices $\RM{T}_{\RV{\theta},\mathrm{S}}$ and $\M{T}_{\RV{\theta},\mathrm{L}}$ are defined as
$$
\begin{aligned}
\RM{T}_{\RV{\theta},\mathrm{S}} &\triangleq \RM{D}_{\RV{\theta},\mathrm{S}}^{-1}\RM{A}_{\RV{\theta},\mathrm{S}}, \\
\M{T}_{\RV{\theta},\mathrm{L}} &\triangleq \M{D}_{\RV{\theta},\mathrm{L}}^{-1}\M{A}_{\RV{\theta},\mathrm{L}}.
\end{aligned}
$$
\begin{IEEEproof} See reference \cite{rai2004the}.
\end{IEEEproof}
\end{lemma}

Let us take a little detour, and consider the following matrix
\begin{equation}
\tilde{\RM{L}}_{\RV{\theta},\mathrm{S}}\triangleq \Big(\M{I}-
\frac{\bar{d}}{\bar{d}+N_\mathrm{p}}\RM{T}_{\RV{\theta},\mathrm{S}}\Big)^{-1}.
\end{equation}
Note that $\frac{\bar{d}}{\bar{d}+N_\mathrm{p}}\in(0,1)$, hence $\frac{1}{N_\mathrm{a}}\mathrm{tr}\{\tilde{\RM{L}}_{\RV{\theta},\mathrm{S}}\}$ corresponds to a Lipschitz function with respect to the eigenvalues of $\RM{T}_{\RV{\theta},\mathrm{S}}$ taking the following form
\begin{equation}
\frac{1}{N_\mathrm{a}}\mathrm{tr}\left\{\tilde{\RM{L}}_{\RV{\theta},\mathrm{S}}\right\} = \int f(\lambda)\mathrm{d} \mu\left(\RM{T}_{\RV{\theta},\mathrm{S}}\right)
\end{equation}
where $f(\lambda) \triangleq \big(1-\frac{\bar{d}}{\bar{d}+N_\mathrm{p}}\cdot \lambda\big)^{-1}$ with Lipschitz constant $\epsilon_L = \frac{\frac{\bar{d}}{\bar{d}+N_\mathrm{ p}}}{\left(1-\frac{\bar{d}}{\bar{d}+N_\mathrm{p}}\right)^2}$ since $\RM{T}_{\RV{\theta},\mathrm{S}}$ has maximum eigenvalue $1$. As long as the inverse of $\tilde{\RM{L}}_{\RV{\theta},\mathrm{ S}}$ exists, we have $\epsilon_L<\infty$. Therefore, $\frac{1}{N_\mathrm{a}}\mathrm{tr}\{\tilde{\RM{L}}_{\RV{\theta},\mathrm{S}}\}$ converges to $\frac{1}{N_\mathrm{a}}\mathrm{tr}\{\tilde{\M{L}}_{\RV{\theta},\mathrm{L}}\}$ corresponding to the lattice network in the manner described in \eqref{concentrate}, since
\begin{equation}
\begin{aligned}
\tilde{\M{L}}_{\RV{\theta},\mathrm{L}} &\triangleq \Big(\M{I}-\frac{\bar{d}}{\bar{d}+N_\mathrm{p}}\M{T}_{\RV{\theta},\mathrm{L}}\Big)^{-1}\\
\frac{1}{N_\mathrm{a}}\mathrm{tr}\left\{\tilde{\M{L}}_{\RV{\theta},\mathrm{L}}\right\} &= \int f(\lambda)\mathrm{d} \mu\left(\M{T}_{\RV{\theta},\mathrm{L}}\right)
\end{aligned}
\end{equation}
as long as the two networks share a common $\bar{d}$ (although $\bar{d}$ has different physical meaning in these networks).

Now the remaining problem is to show the convergence of $\RM{L}_{\RV{\theta},\mathrm{S}}$ to $\tilde{\RM{L}}_{\RV{\theta},\mathrm{S}}$ and that of $\M{L}_{\RV{\theta},\mathrm{L}}$ to $\tilde{\M{L}}_{\RV{\theta},\mathrm{L}}$, in which the former is given by the following lemma.

\begin{lemma}\label{lemma:convergence1}
For $R_\mathrm{max} = \Theta(B^{\frac{1}{k}})$ where $k>2$, there exists positive $t$ such that the following convergence holds
\begin{equation}\label{convergence}
\begin{aligned}
\mathbb{P}\bigg\{\Big|\frac{1}{N_\mathrm{a}}\!\left(\mathrm{tr}\{\RM{L}_{\RV{\theta},\mathrm{S}}\} \!-\! \{\tilde{\RM{L}}_{\RV{\theta},\mathrm{S}}\}\right)\Big|&\!>\!\frac{4t(\log \bar{d})^{\frac{1}{4}}}{\bar{d}}\bigg\} \leq o\bigg(\frac{\bar{d}}{N_\mathrm{a}^3}\bigg).
\end{aligned}
\end{equation}
\begin{IEEEproof} The matrix $\RM{L}_{\RV{\theta},\mathrm{S}}$ can be rewritten as
\begin{equation}
\RM{L}_{\RV{\theta},\mathrm{S}} = \left(\M{I}-\left(\RM{D}_{\RV{\theta},\mathrm{S}}+N_\mathrm{p}\M{I}\right)^{-1}
\RM{A}_{\RV{\theta},\mathrm{S}}\right)^{-1}.
\end{equation}
Now consider a specific agent in the network, say, agent $i$. If the position of the agent, $\V{p}_i$, is given, then the conditional expectation of the number of neighboring agents is given as
\begin{equation}
\tilde{\rv{d}}_i(\V{p}_i)\triangleq\E{d_{{\rm A},i}|\V{p}_i} = \sum_{j\in\Set{A}\backslash\{i\}} \rv{z}_j
\end{equation}
where $\rv{z}_j$ is a random variable taking value of $1$ when agent $j$ is in the communication range of agent $i$ and $0$ otherwise.
Using Lemma 2.1 in \cite{boyd2005mixing}, we have that for $R_\mathrm{max} = \Theta\big(B^{\frac{1}{k}}\big)$ with $k>1$,
\begin{equation}\label{order_d}
\RM{D}_{\RV{\theta},\mathrm{S}} = \diag{\tilde{\rv{d}}_1(\V{p}_1),\tilde{\rv{d}}_1(\V{p}_2),\dotsc,\tilde{\rv{d}}_{N_\mathrm{a}}(\V{p}_{N_\mathrm{a}})} + \RM{E}
\end{equation}
with probability at least $1-\frac{2}{N_\mathrm{a}^3}$, where $\RM{E}$ is a diagonal matrix with the $i$th diagonal element (denoted as $\rv{e}_i$) scales as
$$
\rv{e}_i = \Omega\Big(\sqrt{\tilde{\rv{d}}_i(\V{p}_i)\log \tilde{\rv{d}}_i(\V{p}_i)}\Big).
$$
Since $\M{D}_{\RV{\theta}}^\mathrm{C}$ is a diagonal matrix, with previous assumptions, we can derive \eqref{converge_sto_sto} from \eqref{order_d},
where \eqref{conv_wp} holds with probability at least $1-\frac{2}{N_{\rm a}^3}$, $t_1$ and $t_2$ are constants, $\rv{n}_{\rm int}$ is the number of agents satisfying $\|\bm{p}_i\|_{\infty}\le B-R_{\rm max}$.

Note that $\rv{n}_{\rm int}$ is a sum of random variables taking the following form
\setcounter{equation}{92}
\begin{equation}
\rv{n}_{\rm int} = \sum_{i=1}^{N_{\rm a}} \tilde{\rv{z}}_i
\end{equation}
where $\tilde{\rv{z}}_i$ is a random variable taking value of $1$ when $\|\bm{p}_i\|_{\infty}\le B-R_{\rm max}$ and $0$ otherwise. $\tilde{\rv{z}}_i$'s are i.i.d. Bernoulii variables with $\Prob{\tilde{\rv{z}}_i=1} = \frac{\lambda_{\rm s}(B-R_{\rm max})^2}{N_{\rm a}}$. Thus using the Chernoff bound, we have

\begin{equation}
\begin{aligned}
\!\!\!\!\mathbb{P}\{|\rv{n}_{\rm int}\!-\!\lambda_{\rm s}(B\!-\!R_{\rm max})^2|&\!\ge\!\delta \lambda_{\rm s}(B-R_{\rm max})^2\} \\
&\!\le\! 2\exp\left(\!-\frac{\delta^2\lambda_{\rm s}(B\!-\!R_{\rm max})^2}{2}\right)\!.\!\!
\end{aligned}
\end{equation}

Choosing $\delta = \sqrt{\frac{6\ln N_{\rm a}}{\lambda_{\rm s}(B-R_{\rm max})^2}}$, with probability at least $1-\frac{2}{N_{\rm a}^3}$, we have
\begin{equation}
\left|\rv{n}_{\rm int}-\lambda_{\rm s}(B-R_{\rm max})^2\right|\le \sqrt{6\ln N_{\rm a} \lambda_{\rm s}(B-R_{\rm max})^2}.
\end{equation}
Therefore, with probability at least $1-\frac{2}{N_{\rm a}^3}$,
\begin{equation}\label{N_int}
\begin{aligned}
\frac{N_\mathrm{a}-\rv{n}_\mathrm{int}}{\rv{n}_\mathrm{int}} &\leq \frac{1}{\lambda_\mathrm{s}(B-R_\mathrm{max})^2-\sqrt{6\ln N_\mathrm{a} \lambda_\mathrm{s}(B-R_\mathrm{max})^2}} \\
&\hspace{5mm}\cdot \left(2\lambda_\mathrm{s}BR_\mathrm{max}-\lambda_\mathrm{s}R_\mathrm{max}^2\right. \\
&\hspace{5mm}\left.+\sqrt{6\ln N_\mathrm{a} \lambda_\mathrm{s}(B-R_\mathrm{max})^2}\right) \\
&=\Theta\left(R_\mathrm{max}B^{-1}\right).
\end{aligned}
\end{equation}
Since $R_\mathrm{max} = \Theta(B^{\frac{1}{k}})$ with $k>2$, from \eqref{N_int} we see that $\frac{N_\mathrm{a}-\rv{n}_\mathrm{int}}{\rv{n}_\mathrm{int}} = o(\bar{d}^{-\frac{1}{2}})$. Hence \eqref{conv_wp} can be further rewritten as
\begin{equation}\label{conv_sto_sto_hs}
\big\|\RM{L}_{\RV{\theta},\mathrm{S}}^{-1} - \tilde{\RM{L}}_{\RV{\theta},\mathrm{S}}^{-1}\big\|_\mathrm{HS} \leq t^2\bar{d}^{-2}\sqrt{\log \bar{d}}
\end{equation}
with probability at least $1-\frac{4}{N_\mathrm{a}^3}$, where $t$ is a constant.

Furthermore, given that both $\RM{L}_{\RV{\theta},\mathrm{S}}$ and $\tilde{\RM{L}}_{\RV{\theta},\mathrm{S}}$ are invertible, we can apply the spectral concentration technique used in the derivation of Lemma 2.5 in \cite{rai2004the}, and obtain
\begin{equation}
\begin{aligned}
\!\mathbb{P}\bigg\{\!\Big|\frac{1}{N_\mathrm{a}}\mathrm{tr}\{\RM{L}_{\RV{\theta},\mathrm{S}}\} \!-\! \frac{1}{N_\mathrm{a}}\mathrm{tr}\{\tilde{\RM{L}}_{\RV{\theta},\mathrm{ S}}\}\Big|&\!>\!4\epsilon\!\bigg\} \\
&\!\le\! \frac{4}{\epsilon} \mathbb{P}\Big\{\big\|\RM{L}_{\RV{\theta},\mathrm{S}}^{-1}\!-\!\tilde{\RM{L}}_{\RV{\theta},\mathrm{S}}^{-1}\big\|_\mathrm{HS}\!>\! \epsilon^2\Big\}\!.
\end{aligned}
\end{equation}

Letting $\epsilon = \frac{t\left(\log \bar{d}\right)^{\frac{1}{4}}}{\bar{d}}$, we obtain \eqref{convergence}, thus complete the proof.
\end{IEEEproof}
\end{lemma}

Using similar techniques as applied in the derivation of Lemma \ref{lemma:convergence1}, the following result on the convergence of $\M{L}_{\RV{\theta},\mathrm{L}}$ to $\tilde{\M{L}}_{\RV{\theta},\mathrm{L}}$ can also be obtained.
\begin{corollary}
For $R_\mathrm{max} = \Theta(B^{\frac{1}{k}})$ where $k>2$, we have $\big\|\M{L}_{\RV{\theta},\mathrm{L}}^{-1} - \tilde{\M{L}}_{\RV{\theta},\mathrm{L}}^{-1}\big\|_\mathrm{HS} \leq t_1^2 \bar{d}^{-2}$, and hence
\begin{equation}\label{conv_lattice_lattice_tr}
\Big|\frac{1}{N_\mathrm{a}}\mathrm{tr}\big\{\M{L}_{\RV{\theta},\mathrm{L}}\big\} - \frac{1}{N_\mathrm{a}}\mathrm{tr}\big\{\tilde{\M{L}}_{\RV{\theta},\mathrm{L}}\big\}\Big| \leq \frac{t_2^2}{\bar{d}}
\end{equation}
where $t_1$ and $t_2$ are constants.
\end{corollary}

Finally, with \eqref{delta_ii_approx}, \eqref{concentrate}, \eqref{convergence} and \eqref{conv_lattice_lattice_tr}, by application of the union bound, we obtain \eqref{sto_conv} after some algebra.
\end{IEEEproof}

\bibliographystyle{IEEEtran}
\bibliography{IEEEabrv,Net-Sync,BiblioCV,WGroup}

\begin{IEEEbiography}[{\includegraphics[width=1in,height=1.25in,clip,keepaspectratio]{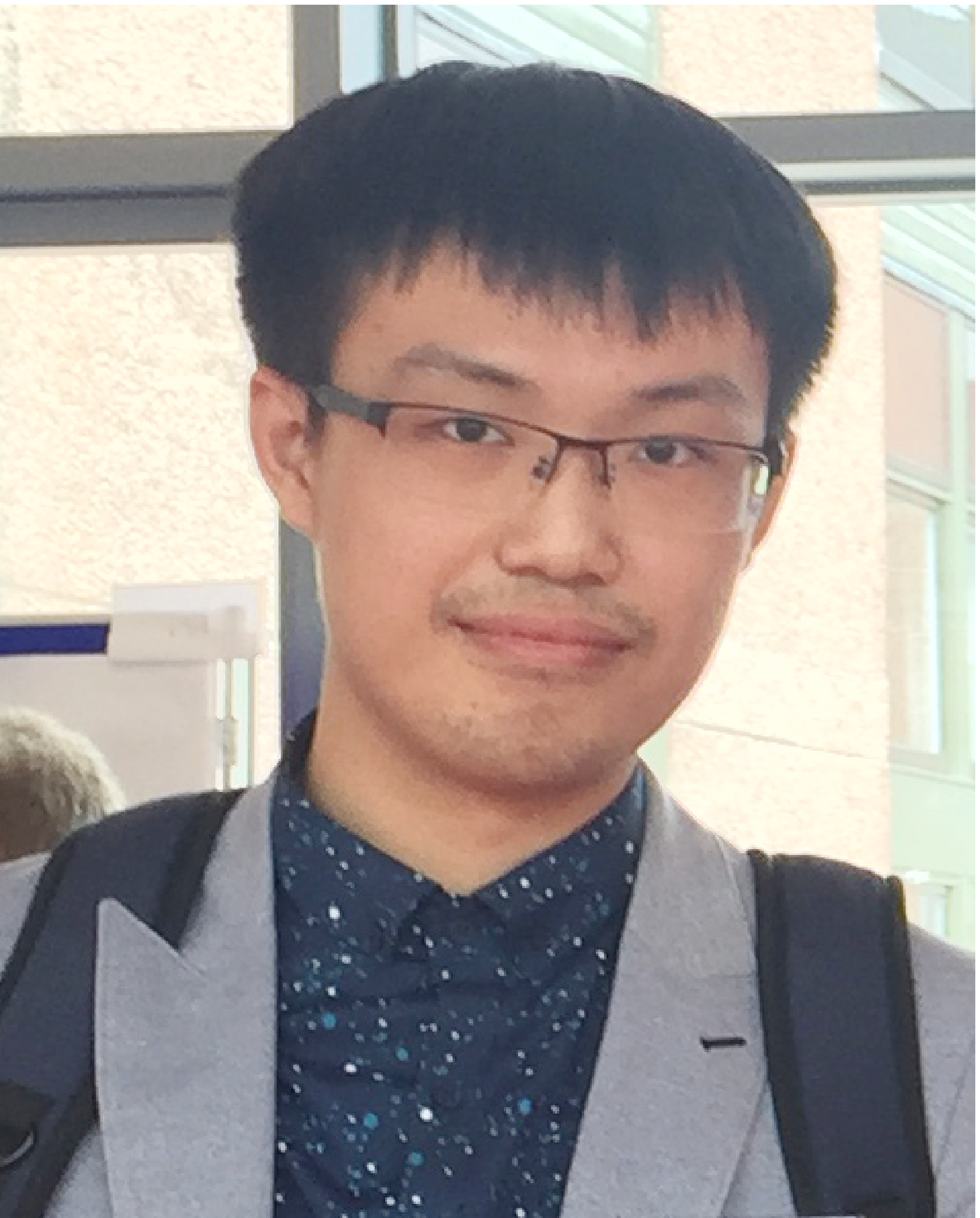}}]{Yifeng Xiong}(S'15) received his B.S. degree from Beijing Institute of Technology (BIT), Beijing, China in 2015. He is currently pursuing the M.S. degree in the School of Information and Electronics at BIT. His research interests include graph theory, statistical inference, and their application to localization and wireless communications. His current research focuses on network localization, network synchronization, and signal processing on graphs. He serves as a reviewer for \textsc{IEEE Communications Letters} and \textsc{IEEE Access}.
\end{IEEEbiography}

\begin{IEEEbiography}[{\includegraphics[width=1in,height=1.25in,clip,keepaspectratio]{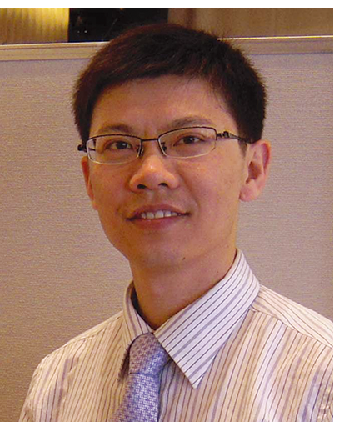}}]{Nan Wu}(M'11) received his B.S., M.S. and Ph.D. degrees from Beijing Institute of Technology (BIT), Beijing, China in 2003, 2005 and 2011, respectively. He is currently an Associate Professor with School of Information and Electronics, BIT. From Nov. 2008 to Nov. 2009, he was a visiting Ph.D. student with Department of Electrical Engineering, Pennsylvania State University, USA. He is the recipient of National Excellent Doctoral Dissertation Award by MOE of China in 2013. His research interests include signal processing in wireless communication networks. He serves as an editorial board member for \textsc{IEEE Access}, \textit{International Journal of Electronics and Communications}, \textit{KSII Transactions on Internet and Information Systems} and \textit{IEICE Transactions on Communications}.
\end{IEEEbiography}

\begin{IEEEbiography}[{\includegraphics[width=1in,height=1.25in,clip,keepaspectratio]{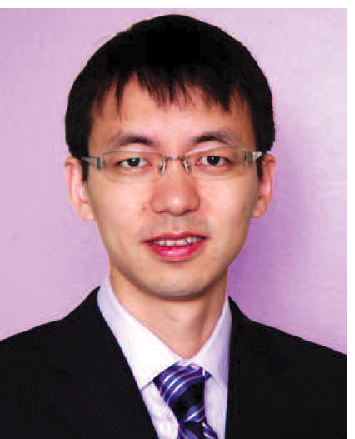}}]{Yuan Shen}(S'05--M'14) received the Ph.D. degree and the S.M. degree in electrical engineering and computer science from the Massachusetts Institute of Technology (MIT), Cambridge, MA, USA, in 2014 and 2008, respectively, and the B.E. degree (with highest honor) in electronic engineering from Tsinghua University, Beijing, China, in 2005. He is an Associate Professor with the Department of Electronic Engineering at Tsinghua University. Prior to that, he was a Research Assistant and then Postdoctoral Associate with the Wireless Information and Network Sciences Laboratory at MIT in 2005-2014. He was with the Hewlett-Packard Labs in winter 2009 and the Corporate R\&D at Qualcomm Inc. in summer 2008. His research interests include statistical inference, network science, communication theory, information theory, and optimization. His current research focuses on network localization and navigation, inference techniques, resource allocation, and intrinsic wireless secrecy.

Prof. Shen was a recipient of the Qiu Shi Outstanding Young Scholar Award (2015), the China's Youth 1000-Talent Program (2014), the Marconi Society Paul Baran Young Scholar Award (2010), and the MIT Walter A. Rosenblith Presidential Fellowship (2005). His papers received the IEEE Communications Society Fred W. Ellersick Prize (2012) and three Best Paper Awards from the IEEE Globecom (2011), ICUWB (2011), and WCNC (2007). He is elected Vice Chair (2017--2018) and Secretary (2015--2016) for the IEEE ComSoc Radio Communications Committee. He serves as TPC symposium Co-Chair for the IEEE Globecom (2016), the European Signal Processing Conference (EUSIPCO) (2016), and the IEEE ICC Advanced Network Localization and Navigation (ANLN) Workshop (2016 and 2017). He also serves as Editor for the \textsc{IEEE Communications Letters} since 2015 and Guest-Editor for the \textit{International Journal of Distributed Sensor Networks} (2015).
\end{IEEEbiography}

\begin{IEEEbiography}[{\includegraphics[width=1in,height=1.25in,clip,keepaspectratio]{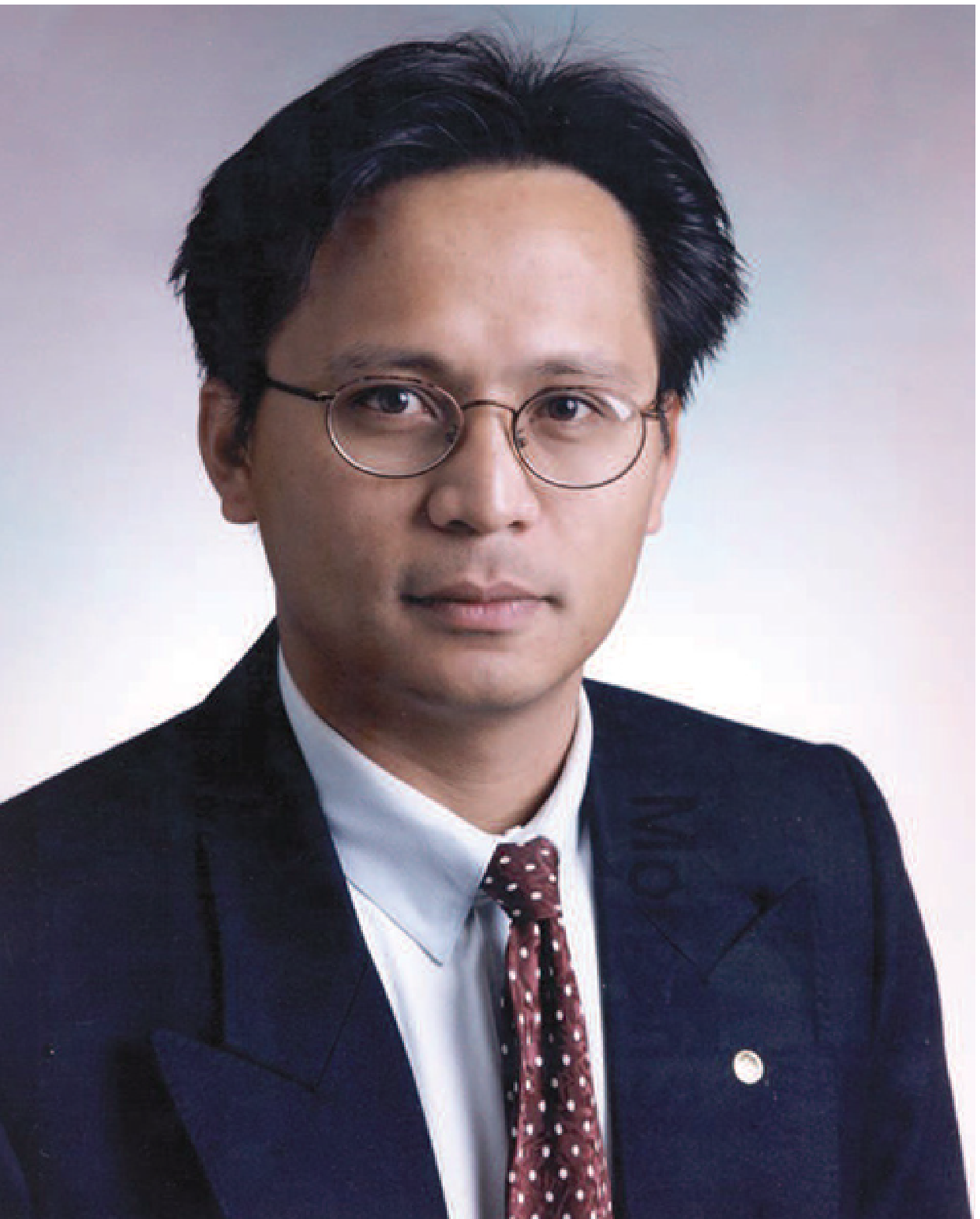}}]{Moe Z. Win} (S'85--M'87--SM'97--F'04) received both the Ph.D. in Electrical Engineering and the M.S. in Applied Mathematics as a Presidential Fellow at the University of Southern California (USC) in 1998. He received the M.S. in Electrical Engineering from USC in 1989 and the B.S. (\emph{magna cum laude}) in Electrical Engineering from Texas A\&M University in 1987. He is a Professor at the Massachusetts Institute of Technology (MIT) and the founding director of the Wireless Information and Network Sciences Laboratory. Prior to joining MIT, he was with AT\&T Research Laboratories for five years and with the Jet Propulsion Laboratory for seven years. His research encompasses fundamental theories, algorithm design, and experimentation for a broad range of real-world problems. His current research topics include network localization and navigation, network interference exploitation, intrinsic wireless secrecy, adaptive diversity techniques, and ultra-wideband systems.

Professor Win is an elected Fellow of the AAAS, the IEEE, and the IET, and was an IEEE Distinguished Lecturer. He was honored with two IEEE Technical Field Awards: the IEEE Kiyo Tomiyasu Award (2011) and the IEEE Eric E. Sumner Award (2006, jointly with R. A. Scholtz). Together with students and colleagues, his papers have received numerous awards, including the IEEE Communications Society's Stephen O. Rice Prize (2012), the IEEE Aerospace and Electronic Systems Society's M. Barry Carlton Award (2011), the IEEE Communications Society's Guglielmo Marconi Prize Paper Award (2008), and the IEEE Antennas and Propagation Society's Sergei A. Schelkunoff Transactions Prize Paper Award (2003). Highlights of his international scholarly initiatives are the Copernicus Fellowship (2011), the Royal Academy of Engineering Distinguished Visiting Fellowship (2009), and the Fulbright Fellowship (2004). Other recognitions include the IEEE
Communications Society Edwin H. Armstrong Achievement Award (2016), the International Prize for Communications Cristoforo Colombo (2013), the Laurea Honoris Causa from the University of Ferrara (2008), the Technical Recognition Award of the IEEE ComSoc Radio Communications Committee (2008), and the U.S. Presidential Early Career Award for Scientists and Engineers (2004).

Dr. Win was an elected Member-at-Large on the IEEE Communications Society Board of Governors (2011--2013). He was the Chair (2005--2006) and Secretary (2003--2004) for the Radio Communications Committee of the IEEE Communications Society. Over the last decade, he has organized and chaired numerous international conferences. He is currently serving on the SIAM Diversity Advisory Committee and the advisory board of the \textsc{IEEE Communications Letters}. He served as Editor-at-Large (2012--2015) for the \textsc{IEEE Wireless Communications Letters}, as Editor (2006--2012) for the \textsc{IEEE Transactions on Wireless Communications}, and as Area Editor (2003--2006) and Editor (1998--2006) for the \textsc{IEEE Transactions on Communications}. He was Guest-Editor for the \textsc{Proceedings of the IEEE} (2009) and for the \textsc{IEEE Journal on Selected Areas in Communications} (2002).
\end{IEEEbiography}
\end{document}